\newcommand{\be}{\begin{equation}}
\newcommand{\ee}{\end{equation}}
\newcommand{\bea}{\begin{eqnarray}}
\newcommand{\eea}{\end{eqnarray}}
\newcommand{\nn}{\nonumber \\}
\newcommand{\p}[1]{(\ref{#1})}
\newcommand{\bQ}{{\overline Q}}
\newcommand{\bD}{{\overline D}}
\newcommand{\tV}{{\widetilde V}}
\newcommand{\cL}{{\cal L}}
\newcommand{\cZ}{{\cal Z}}
\newcommand{\cD}{{\cal D}}
\newcommand{\hK}{{\hat K}}
\newcommand{\hD}{{\hat D}}
\newcommand{\cV}{{\cal V}}
\newcommand{\bV}{{\overline V}}
\newcommand{\bLam}{{\overline \Lambda}}
\newcommand{\bT}{{\overline T}}
\newcommand{\bS}{{\overline S}}
\newcommand{\cU}{{\cal U}}
\newcommand{\cQ}{{\cal Q}}
\newcommand{\cW}{{\cal W}}

\newcommand{\cbZ}{\overline{\cal Z}}
\newcommand{\cbV}{\overline{\cal V}}
\newcommand{\cbD}{\overline{\cal D}}
\newcommand{\bF}{{\overline F}}
\newcommand {\fac}{{(1 + \alpha^2 z \overline{z})}}
\newcommand{\bz}{{\bar z}}
\newcommand{\bt}{{\bar\theta}}
\newcommand{\bomega}{{\bar\omega}}
\newcommand{\beps}{{\bar\epsilon }}

\newcommand{\bpsi}{{\bar\psi}}
\newcommand {\fc}{{1 + \alpha^2 z {\overline{z}}}}
\newcommand{\vt}{\vartheta}
\documentclass[vecphys]{svmult}
\usepackage{amscd,amsmath,amssymb}

\usepackage{makeidx}         
\usepackage{graphicx}        
\usepackage{multicol}        

\makeindex             

\begin{document}

\title*{Supersymmetric Mechanics in Superspace}
\author{Stefano Bellucci\inst{1}\and Sergey Krivonos\inst{2}}
\authorrunning{S.~Bellucci and S.~Krivonos}
\institute{INFN-Laboratori Nazionali di Frascati, \\
 C.P. 13, 00044 Frascati, Italy \and
Bogoliubov Laboratory of Theoretical Physics, \\
Joint Institute for Nuclear Research, \\
141980, Dubna, Moscow region, Russia}


%
%
\maketitle

\section{Introduction}\label{sec:0}
These Lectures have been given at Laboratori Nazionali di Frascati in the month of March, 2005.
The main idea was to provide our young collegues, who joined us in our attempts   to understand
the structure of $N$-extended supersymmetric one-dimensional systems, with short
descriptions of the methods and techniques we use. This was reflected in the choice of
material and in the style of presentation. We base our treatment mainly on the superfield point of view.
Moreover, we prefer to deal with $N=4$ and $N=8$ superfields. At present, the exists an
extensive literature on the components approach to extended supersymmetric theories in $d=1$
while the manifestly supersymmetric formulation in terms of properly constrained superfields
is much less known. Nevertheless,  we believe that just such formulations are preferable.

In order to make these Lectures more or less self-consistent, we started from the simplest
examples of one-dimensional supersymmetric theories and paid a lot of attention to the
peculiarities of $d=1$ supersymmetry. From time to time we presented the calculations
in a very detailed way. In other cases we omitted the details and gave only the final answers.
In any case these Lectures cannot be considered as a textbook in any respect. They can be
considered as our personal point of view on the one-dimensional superfield theories and on the
methods and techniques we believe to be important.

Especially, all this concerns Section 3, were we discuss the nonlinear realizations
method. We did not present any proofs in this Section. Instead we focused on the details of calculations.

Finally, we apologize for the absolutely incomplete list of References.

\section{Supersymmetry in $d=1$}\label{sec:1}
The extended supersymmetry in one dimension has plenty of peculiarities which
make it quite different from higher-dimensional analogs. Indeed, even the basic
statement of any supersymmetric theory in $d>1$ -- the equality of bosonic and
fermionic degrees of freedom -- is not valid in $d=1$. As a result, there are many
new supermultiplets in one dimension which have no higher-dimensional ancestors.
On the other hand, the constraints describing on-shell supermultiplets, being
reduced to $d=1$, define off-shell multiplets! Therefore, it makes sense to
start with the basic properties of one dimensional supermultiplets and give
a sort of vocabulary with all linear finite-dimensional $N=1,2,4$ and $N=8$
supermultiplets. This is the goal of the present Section.

\subsection{Super-Poincar\'e algebra in $d=1$}\label{subsec:11}
In one dimension there is no Lorentz group and therefore all bosonic
and fermionic fields have no space-time indices. The simplest free action
for one bosonic field $\phi$ and one fermionic field $\psi$ reads
\begin{equation} \label{eq1}
S= \gamma \int dt \left[ {\dot\phi}^2 - \frac{i}{2} \psi\dot\psi \right].
\end{equation}
In what follows it will be useful to treat the scalar field as dimensionless and
assign dimension $cm^{-1/2}$ to fermions. Therefore, all our actions will contain
the parameter $\gamma$ with the dimension $[\gamma ]=cm$.

The action \p{eq1} provides the first example of a supersymmetric invariant action.
Indeed, it is a rather simple exercise to check its invariance with respect to the following
transformations:
\be\label{eq2}
\delta \phi = -i \epsilon \psi,\quad \delta \psi = - \epsilon \dot\phi \;.
\ee
As usual, the infinitesimal parameter $\epsilon$ anticommutes with fermionic fields
and with itself. What is really important about transformations \p{eq2} is their commutator
\bea\label{eq3}
\delta_2 \delta_1 \phi =\delta_2\left( -\epsilon_1 \psi\right) =i \epsilon_1\epsilon_2 \dot\phi,&&\nn
\delta_1 \delta_2 \phi = i  \epsilon_2\epsilon_1 \dot\phi &&\Rightarrow
\left[ \delta_2,\delta_1\right] \phi =2i \epsilon_1\epsilon_2 \dot\phi \;.
\eea
Thus, from \p{eq3} we may see the main property of supersymmetry transformations: they commute
on translations. In our simplest one-dimensional framework this is the time translation. This
property has the following form in terms of the supersymmetry generator $Q$:
\be\label{eq4}
\left\{ Q,Q \right\}= -2P .
\ee
The anticommutator \p{eq4}, together with
\be
\left[ Q, P\right] =0
\ee
describe $N=1$ super-Poincar\'e algebra in $d=1$. It is rather easy to guess the structure of
$N$-extended super-Poincar\'e algebra: it includes $N$ {\em real} supercharges $Q^A, A=1,...,N$
with the following commutators:
\be\label{eq6}
\left\{ Q^A,Q^B \right\}= -2\delta^{AB} P ,\; \left[ Q^A, P\right] =0 \;.
\ee
Let us stress that the reality of the supercharges is very important, as well as having the same
sign in the r.h.s. of $\left\{ Q^A,Q^B \right\}$ for all $Q^A$. From time to time one can see
in the literature wrong statements about the number of supersymmetries in the theories when authors
forget about these absolutely needed properties.

{}From \p{eq2} we see that the minimal $N=1$ supermultiplet includes one bosonic and one
fermionic field. A natural question arises: how many components we need, in order to realize the $N$-extended
superalgebra \p{eq6}? The answer has been found in a paper by S.J.Gates and L.Rana \cite{GR}.
Their idea is to mimic the transformations \p{eq2} for all $N$ super-translations as follows:
\be\label{eq7}
\delta \phi_i = -i \epsilon^A \left( L_A \right)_i^{\hat i} \psi_{\hat i},\quad
\delta \psi_{\hat i} = - \epsilon^A \left( R_A \right)_{\hat i}^i \dot\phi_i \;.
\ee
Here the indices $i=1,\ldots,d_b$ and ${\hat i}=1,\ldots,d_f$ count the numbers of bosonic
and fermionic components, while $\left( L_A \right)_i^{\hat i}$ and $\left( R_A \right)_{\hat i}^i$
are $N$ arbitrary, for the time being, matrices. The additional conditions one should impose on the transformations
\p{eq7} are
\begin{itemize}
\item They should form the $N$-extended superalgebra \p{eq6}
\item They should leave invariant the free action constructed from the involved fields.
\end{itemize}
These conditions result in some equations on the matrices $L_A$ and $R_A$ which has been solved in \cite{GR}.
These results for the most interesting cases are summarized in the Table \p{tab:1}.
\begin{table}
\centering
\caption{Minimal supermultiplets in $N$-extended supersymmetry}
\label{tab:1}
\begin{tabular}{|c|c|c|c|c|c|c|c|c|c|c|c|c|}
\hline\noalign{\smallskip}
N & 1 & 2& 3& 4& 5& 6&7&8& 9& 10& 12& 16   \\
\noalign{\smallskip}\hline\noalign{\smallskip}
d & 1 & 2& 4& 4& 8& 8&8&8& 16& 32& 64& 128   \\
\noalign{\smallskip}\hline
\end{tabular}
\end{table}
Here, $d=d_b=d_f$ is the number of bosonic/fermionic components. From Table \p{tab:1}
we see that there are four special cases with $N=1,2,4,8$ when $d$ coincides with $N$. Just
these cases we will discuss in the present Lectures. When $N>8$ the minimal dimension of the supermultiplets
rapidly increases and the analysis of the corresponding theories becomes very complicated.
{}For many reasons, the most interesting case seems to be the ${\cal
N}=8$ supersymmetric mechanics. Being the highest $\cal N$
case of {\it minimal} ${\cal N}$-extended supersymmetric mechanics
admitting realization on ${\cal N}$ bosons (physical and
auxiliary) and ${\cal N}$ fermions, the  systems with eight
supercharges are the highest $\cal N$ ones, among the extended
supersymmetric systems, which still possess a non-trivial geometry
in the bosonic sector \cite{VP1}. When the number of supercharges
exceeds 8, the target spaces are restricted to be symmetric
spaces. Moreover, ${\cal N}=8$ supersymmetric mechanics   should be related via
a proper dimensional reduction with four-dimensional  ${\cal N}=2$
supersymmetric field theories. So, one may hope that some
interesting properties of the latter will survive after reduction.

\subsection{Auxiliary fields}
In the previous Subsection we considered the realization of $N$-extended
supersymmetry on the bosonic fields of the same dimensions. Indeed, in \p{eq7}
all fields appear on the same footing. It is a rather special case which occurs
only in $d=1$. In higher dimensions the appearance of the auxiliary fields is
inevitable. They appear also in one dimension. Moreover, in $d=1$ one
may convert any physical field to auxiliary and vice versa \cite{GR}. In order to
clarify this very important property of one dimensional theories, let us
consider the simplest example of $N=2,d=1$ supermultiplets.

The standard definition of $N=2, d=1$ super-Poincar\'e algebra follows from \p{eq6}:
\be\label{21}
\left\{ Q^1,Q^1 \right\}=\left\{ Q^2,Q^2 \right\}= -2 P ,\;
\left\{ Q^1,Q^2 \right\}=0.
\ee
It is very convenient to redefine the supercharges as follows:
\bea\label{22}
Q \equiv \frac{1}{\sqrt{2}} \left( Q_1+iQ_2 \right),\;
\bQ \equiv \frac{1}{\sqrt{2}} \left( Q_1-iQ_2 \right), \nn
\left\{ Q,\bQ \right\}= -2 P ,\; Q^2 =\bQ{}^2=0.
\eea
{}From Table 1 we know that the minimal $N=2$ supermultiplet contains
two bosonic and two fermionic components. The supersymmetry transformations
may be easily written as
\be\label{23}
\left\{ \begin{array}{l}
\delta \phi =-i \bar\epsilon \psi, \\
\delta \psi= -2 \epsilon \dot\phi,
\end{array} \right. \quad
\left\{ \begin{array}{l}
\delta \bar\phi =-i \epsilon \bar\psi, \\
\delta \bar\psi= -2 \bar\epsilon \dot{\bar\phi}.
\end{array} \right. \quad
\ee
Here, $\phi$ and $\psi$ are complex bosonic and fermionic components.
The relevant free action has a very simple form
\be\label{24}
S= \gamma \int dt \left( \dot{\phi} {\dot{\bar\phi}}- \frac{i}{2} \psi {\dot{\bar\psi}} \right).
\ee

Now we introduce the new bosonic variables $V$ and $A$
\be\label{25}
V= \phi+\bar\phi,\; A= i\left( \dot{\phi} -\dot{\bar{\phi}} \right).
\ee
What is really impressive is that, despite the definition of $A$ in terms of time derivatives of
the initial bosonic fields, the supersymmetry transformations can be written in terms of
$\{V,\psi,\bar\psi,A\}$ only
\bea\label{26}
\delta V=-i\left(\bar\epsilon \psi+\epsilon\bar\psi \right),\; \delta A= \bar\epsilon\dot{\psi} -\epsilon\dot{\bar\psi},\nn
\delta\psi=-\epsilon \left( \dot{V} -iA\right),\; \delta\bar\psi =-\bar\epsilon\left( \dot{V}+iA\right).
\eea
Thus, the fields $V,\psi,\bar\psi,A$ form a $N=2$ supermultiplet, but the dimension of the component $A$ is now $cm^{-1}$.
If we rewrite the action \p{24} in terms of new components
\be\label{27}
S= \gamma \int dt \left(\frac{1}{4} \dot{V} \dot{V} - \frac{i}{2} \psi \dot{\bar\psi} +\frac{1}{4}A^2\right),
\ee
one may see that the field $A$ appears in the action without derivatives and their equation of motion
is purely algebraic
\be\label{28}
A=0 \;.
\ee
In principle, we can exclude this component from the Lagrangian \p{27} using \p{28}. As a result we will have the Lagrangian
written in terms of $V,\psi,\bar\psi$ only. But, without $A$, supersymmetry will close only up to equations of motion.
Indeed, the variation of \p{28} with respect to \p{26} will enforce equations of motion for fermions
$$ \dot\psi=\dot{\bar\psi}=0.$$
The next variation of these equations give
$$ \ddot{V}=0.$$
Such components are called auxiliary fields. In what follows we will use the notation
${\bf (n,N,N-n)}$ to describe a supermultiplet with $n$ physical bosons, $N$ fermions and $N-n$ auxiliary bosons.
Thus, the transformations \p{26} describe passing from the multiplet ${\bf (2,2,0)}$ to the ${\bf (1,2,1)}$ one.
One may continue this process and so pass from the multiplet ${\bf (1,2,1)}$ to the ${\bf (0,2,2)}$ one, by
introducing the new components $B=\dot{\phi}, \bar{B}=\dot{\bar\phi}$. The existence of such a multiplet containing no physical
bosons at all is completely impossible in higher dimensions. Let us note that the inverse procedure is
also possible \cite{GR}. Therefore, the field contents of {\em linear, finite dimensional} off-shell multiplets
of $N=2,4,8$ $d=1$ supersymmetry read
\bea\label{29}
&& N=2: \; {\bf (2,2,0)},\;{\bf (1,2,1)}\;{\bf (0,2,2)} \nn
&& N=4: \; {\bf (4,4,0)},\;{\bf (3,4,1)},\;{\bf (2,4,2)},\;{\bf (1,4,3)},\;{\bf (0,4,4)} \nn
&& N=8: \; {\bf (8,8,0)},\; \ldots ,\;{\bf (0,8,8)}
\eea
Finally, one should stress that both restrictions -- linearity and finiteness -- are important.
There are nonlinear supermultiplets \cite{HP,ikl1,bikl1}, but it is not always possible to change
their number of physical/auxiliary components. In the case of $N=8$ supersymmetry one may define
infinite-dimensional supermultiplets, but for them the interchanging of the physical and auxiliary
components is impossible. We notice, in passing, that recently
for the construction of $N=8$ supersymmetric mechanics \cite{bikl2,bikl1,Belln8}
the nonlinear chiral multiplet has been used \cite{8nonlin}.

\subsection{Superfields}
Now we will turn to the main subject of these Lectures -- Superfields in $N$-extended $d=1$
Superspace. One may ask - Why do we need superfields? There are a lot of motivations, but here
we present only two of them. First of all, it follows from the previous Subsections that only
a few supermultiplets from the whole "zoo" of them presented in \p{29}, contain no auxiliary fields.
Therefore, for the rest of the cases, working in terms of physical components we will deal with on-shell
supersymmetry. This makes life very uncomfortable - even checking the supersymmetry invariance of
the action becomes a rather complicated task, while in terms of superfields everything is manifestly
invariant. Secondly, it is a rather hard problem to write the interaction terms in the component
approach. Of course, the superfields approach has its own problems. One of the most serious, when
dealing with extended supersymmetry, is to find the irreducibility constraints which decrease the
number of components in the superfields. Nevertheless, the formulation of the theory in a manifestly
supersymmetric form seems preferable, not only because of its intrinsic beauty, but also since it
provides an efficient technique, in particular in quantum calculations.

The key idea of manifestly invariant formulations of supersymmetric theories is using superspace,
where supersymmetry is realized geometrically by coordinate transformations. Let us start with
$N=2$ supersymmetry. The natural definition of $N=2$ superspace $\mathbb{R}^{(1|2)}$ involves time $t$ and two
anticommuting coordinates $\theta,\bar\theta$
\be\label{31a}
\mathbb{R}^{(1|2)} = \left( t, \theta, \bar\theta \right).
\ee
In this superspace $N=2$ super-Poincar\'e algebra \p{22} can be easily realized
\be\label{32a}
\delta \theta=\epsilon,\; \delta \bar\theta =\bar\epsilon,\;
\delta t = -i\left( \epsilon \bar\theta + \bar\epsilon \theta \right).
\ee
$N=2$ superfields $\Phi(t,\theta,\bar\theta)$ are defined as functions on this superspace. The simplest
superfield is the scalar one, which transforms under \p{32} as follows:
\be\label{33a}
\Phi'(t',\theta',\bar\theta{}')=\Phi(t,\theta,\bar\theta).
\ee
{}From \p{33a} one may easily find the variation of the superfield in passive form
\bea\label{34a}
\delta \Phi &&\equiv \Phi'(t,\theta,\bar\theta)-\Phi(t,\theta,\bar\theta) =
-\epsilon \left( \frac{\partial}{\partial\theta} -i \bar\theta \frac{\partial}{\partial t}\right) \Phi -
-\bar\epsilon \left( \frac{\partial}{\partial \bar\theta} -i \theta \frac{\partial}{\partial t}\right) \Phi  \nn
&&\equiv - \epsilon Q \Phi - \bar\epsilon \bQ \Phi .
\eea
Thus we get the realization of supercharges $Q,\bQ$ in superspace
\be\label{35}
Q=\frac{\partial}{\partial\theta} -i \bar\theta \frac{\partial}{\partial t},\;
\bQ=\frac{\partial}{\partial \bar\theta} -i \theta \frac{\partial}{\partial t},\;
\left\{ Q,\bQ\right\} =-2i \frac{\partial}{\partial t} .
\ee

In order to construct covariant objects in superspace, we have to define covariant derivatives
and covariant differentials of the coordinates. Under the transformations \p{32a}, $d\theta$ and $d\bar\theta$
are invariant, but $dt$ is not. It is not too hard to find the proper covariantization of $dt$
\be\label{36}
dt \;\; \rightarrow \;\; \triangle t = dt -i d\bar\theta \theta -i d\theta \bar\theta \;.
\ee
Indeed, one can check that $\delta\triangle t =0$ under \p{32a}. Having at hands the covariant differentials,
one may define the covariant derivatives
\bea\label{37}
\left( dt  \frac{\partial}{\partial t} + d\theta \frac{\partial}{\partial \theta}+
d\bar\theta\frac{\partial}{\partial \bar\theta}\right) \Phi \equiv \left( \triangle t \nabla_t +d\theta \; D +
 d\bar\theta\; \bD \right) \Phi \;, \nn
 \nabla_t = \frac{\partial}{\partial t},\;
 D=\frac{\partial}{\partial\theta} +i \bar\theta \frac{\partial}{\partial t},\;
\bD=\frac{\partial}{\partial \bar\theta} +i \theta \frac{\partial}{\partial t},\; \left\{ D,\bD\right\}= 2i\partial_t.
\eea
As important properties of the covariant derivatives let us note that they anticommute with
the supercharges \p{25}.

The superfield $\Phi$ contains the ordinary bosonic and fermionic fields as coefficients in its
$\theta,\bar\theta$ expansion. A convenient covariant way to define these components is to define
them as follows:
\be\label{38}
V=\Phi|,\; \psi= i\bD \Phi|,\; \bar\psi =- D\Phi|,\; A=\frac{1}{2} \left[ D, \bD \right]\Phi|,
\ee
where $|$ means the restriction to $\theta=\bar\theta=0$. One may check that the transformations of the
components \p{38}, which follow from \p{34a}, coincide with \p{26}. Thus, the general bosonic
$N=2$ superfield $\Phi$ describes the ${\bf (1,2,1)}$ supermultiplet. The last thing we need to know, in order
to construct the superfield action is the rule for integration over Grassmann coordinates $\theta,\bar\theta$.
By definition, this integration is equivalent to a differentiation
\be\label{39}
\int dt d\theta d\bar\theta \cL \equiv \int dt D \bD \cL= \frac{1}{2} \int dt  \left[ D, \bD \right] \cL .
\ee
Now we are ready to write the free action for the $N=2$ ${\bf (1,2,1)}$ supermultiplet
\be\label{40}
S= \gamma \int dt d\theta d\bar\theta \; D \Phi \bD \Phi \;.
\ee
It is a simple exercise to check that, after integration over $d\theta ,d \bar\theta$ and passing to the
components \p{38}, the action \p{40} coincides with \p{27}.

The obvious question now is how to describe in superfields the supermultiplet ${\bf (2,2,0)}$. The latter contains
two physical bosons, therefore the proper superfield should be a complex one. But without any additional
conditions the complex $N=2$ superfield $\phi,\bar\phi$ describes a ${\bf (2,4,2)}$ supermultiplet. The solution
is to impose the so called chirality constraints on the superfield
\be\label{41}
D \phi =0,\quad \bD \bar\phi=0 \;.
\ee
It is rather easy to find the independent components of the chiral superfield \p{41}
\be\label{42}
\tilde\phi = \phi|,\; \bar\psi = i\bD \phi|,\quad \widetilde{\bar\phi}=\bar\phi|,\; \psi=iD \bar\phi|.
\ee
Let us note that, due to the constraints \p{41}, the auxiliary components are expressed through time derivatives
of the physical ones
\be\label{43}
A= \frac{1}{2} \left[ D, \bD\right] \phi| =\frac{1}{2} D\bD \phi| = i\dot{\tilde\phi}.
\ee
The free action has a very simple form
\be\label{44}
S=\gamma \int dt d\theta d\bar\theta \; \left( \phi \dot{\bar\phi} -\dot\phi {\bar\phi}\right)\;.
\ee

As the immediate result of the superfields formulation, one may write the actions of $N=2$ $\sigma$-models
for both supermultiplets
\bea \label{45}
S_\sigma=\gamma  \int dt d\theta d\bar\theta \; F_1(\Phi) D \Phi \bD \Phi \;, \nn
S_\sigma=\gamma \int dt d\theta d\bar\theta \; F_2(\phi,\bar\phi)\left( \phi \dot{\bar\phi} -\dot\phi {\bar\phi}\right)\;,
\eea
where $F_1(\Phi$ and $F_2(\phi,\bar\phi)$ are two arbitrary functions defining the metric in the target space.

Another interesting example is provided by the action of $N=2$ superconformal mechanics \cite{AkPa}
\be\label{46}
S_{Conf}=\gamma \int dt d\theta d\bar\theta \; \left[ D \Phi \bD \Phi  +2m \log \Phi \right].
\ee

The last $N=2$ supermultiplet in Table \p{29}, with content ${\bf (0,2,2)}$,
may be described by the chiral fermionic superfield $\Psi$:
\be\label{47}
D \Psi =0,\quad \bD \bar\Psi=0 \;.
\ee

Thus we described in superfields all $N=2$ supermultiplets. But really interesting features appear in the $N=4$ supersymmetric
theories which we will consider in the next subsection.

\subsection{N=4 Supermultiplets}
The $N=4, d=1$ superspace $\mathbb{R}^{(1|4)}$ is parameterized by the coordinates
\be\label{51}
\mathbb{R}^{(1|4)} = \left( t, \theta_i, \bar\theta{}^j \right), \; \left( \theta_i \right)^\dagger =\bar\theta{}^i,\; i,j=1,2.
\ee
The covariant derivatives may be defined in a full analogy with the $N=2$ case as
\be\label{52}
D^i=\frac{\partial}{\partial\theta_i} +i \bar\theta{}^i \frac{\partial}{\partial t},\;
\bD_j=\frac{\partial}{\partial \bar\theta{}^j} +i \theta_j \frac{\partial}{\partial t},\;
\left\{ D^i,\bD_j\right\}= 2i\delta^i_j\partial_t.
\ee
Such a representation of the algebra of $N=4, d=1$ spinor covariant derivatives manifests an automorphism $SU(2)$
symmetry  (from the full $SO(4)$ automorphism symmetry of $N=4, d=1$ superspace)  realized on the doublet indices $i,j$.
The transformations from the coset  $SO(4)/SU(2)$ rotate $D^i$ and
$\bD{}^j$ through each other.

Now, we are going to describe in superfields all possible $N=4$ supermultiplets from Table  \p{29}.

\subsubsection{N=4, d=1 ``hypermultiplet'' -- ${\bf (4,4,0)}$. }
We shall start with the most general case when the supermultiplet
contains four physical bosonic components. In order to describe this supermultiplet we have to introduce
four $N=4$ superfields $q^i, {\bar q}_j$. These superfields should be properly constrained to
reduce 32 bosonic and 32 fermionic components, which are present in the unconstrained  $q^i, {\bar q}_j$, to
4 bosonic and 4 fermionic ones. One may show that the needed constraints read
\be\label{hyper4a}
D^{(i}q^{j)}=0,\; \bD{}^{(i}q^{j)}=0,\qquad D^{(i}{\bar q}{}^{j)}=0,\; \bD{}^{(i}{\bar q}{}^{j)}=0.
\ee
This $N=4,d=1$ multiplet was considered in  \cite{{CP},{GPS},{HP},{Hull1},{ikl1}} and also was recently
studied in $N=4, d=1$ harmonic superspace \cite{Olaf}.
It resembles the $N=2, d=4$ hypermultiplet.
However, in contrast to the $d=4$ case, the constraints \p{hyper4} define an off-shell
multiplet in $d=1$.

The constraints \p{hyper4a} leave in the $N=4$ superfield $q^i$ just four spinor components
\be
D_i q^i,\; \bD_i q^i,\; D_i,\; {\bar q}{}^i,\; \bD_i {\bar q}{}^i,
\ee
while all higher components in the $\theta$-expansion are expressed as time-derivatives of
the lowest ones. This can be immediately seen from the following consequences of \p{hyper4a}:
$$
D^i \bD_j q^j = 4i \dot{q}{}^i,\quad D^i D_j q^j =0 \;.
$$
The general sigma-model action for the supermultiplet  ${\bf (4,4,0)}$ reads\footnote{The standard
convention for integration in $N=4,d=1$ superspace is
$\int dt d^4 \theta = \frac{1}{16} \int dt D^i D_i \bD{}^j \bD_j$.}
\be\label{action1}
S_\sigma = \int dt d^4\theta K(q, {\bar q}),
\ee
where $K(q, {\bar q})$ is an arbitrary function on $q^i$ and $q_j$. When expressed in components, the action \p{action1}
has the following form:
\be\label{action1b}
S_\sigma = \int dt\left[ \frac{\partial^2 K(q, {\bar q})}{\partial q^i \partial {\bar q}_i} {\dot q}{}^j\dot{\bar q}_j+
\mbox{fermions}\right].
\ee

Another interesting example is the superconformally invariant superfield action\cite{ikl1}
\be\label{hyper8}
S_{Conf}= - \int dt d^4\theta\;
\frac{\mbox{ ln } \left( q^i {\bar q}_i\right)}{q^j {\bar q}_j} \;.
\ee

A more detailed discussion of possible actions for the $q^i$ multiplet
can be found in \cite{Olaf}.
In particular, there exists a superpotential-type off-shell invariant which,
however, does not give
rise in components to any scalar potential for the physical bosons. Instead,
it produces a Wess-Zumino type term of the first order in the
time derivative. It can be interpreted as a coupling to a
four-dimensional background abelian
gauge field. The superpotential just mentioned admits a concise
manifestly supersymmetric superfield formulation,
as an integral over an analytic subspace of $N=4, d=1$ harmonic superspace
\cite{Olaf}.

Notice that the $q^i$ supermultiplet can be considered as a fundamental
one, since all other $N=4$ supermultiplets can be obtained from $q^i$ by reduction. We will consider
how such reduction works in the last section of this Lectures.

\subsubsection{N=4, d=1 ``tensor'' multiplet -- ${\bf (3,4,1)}$.}

The ``tensor'' multiplet includes three $N=4$ bosonic superfields which can be combined
in a $N=4$ isovector real superfield $V^{ij}$ ($V^{ij}=V^{ji}$ and
$\overline{V^{ik}}=\epsilon_{ii'}\epsilon_{kk'}V^{i'k'}$).
The irreducibility constraints may be written in the manifestly $SU(2)$-symmetric form
\be\label{tensor4}
D^{(i}V^{jk)} =0 \; , \quad \bD{}^{(i}V^{jk)} =0 \; .
\ee
The constraints \p{tensor4} could be obtained
by a direct dimensional reduction from the constraints
defining the $N=2, d=4$ tensor multiplet \cite{tensord4}, in which
one suppresses the $SL(2,C)$ spinor indices of the
$d=4$ spinor derivatives, thus keeping only the doublet indices of
the $R$-symmetry $SU(2)$ group.
This is the reason why we can call it
$N=4, d=1$ ``tensor'' multiplet. Of course,
in one dimension no differential (notoph-type) constraints arise
on the components of the superfield $V^{ij}$. The constraints \p{tensor4}
leave in $V^{ik}$ the following independent superfield projections:
\be\label{tensor5}
 V^{ik}~, \; D^{i}V^{kl} =
-\frac{1}{3}(\epsilon^{ik}\chi^l + \epsilon^{il} \chi^k)~, \;
\bar D^{i}V^{kl} = \frac{1}{3}(\epsilon^{ik}\bar\chi^l
+ \epsilon^{il} \bar\chi^k)~, \;
D^i\bar D^k V_{ik}~,
\ee
where
\be
\chi^k \equiv D^iV_i^k~, \quad \bar\chi_k = \overline{\chi^k} = \bar D_i V^i_k~.
\ee
Thus its off-shell component field content is just ${\bf (3,4,1)}$.
The $N=4, d=1$ superfield
$V^{ik}$ subjected to the conditions \p{tensor4}
was introduced in \cite{CR} and,
later on, rediscovered in \cite{ismi,bepa,stro2}.

As in the case of the superfield $q^i$,
the general sigma-model action for the supermultiplet  ${\bf (3,4,1)}$
may be easily written as
\be\label{action2}
S_\sigma = \int dt d^4\theta K(V).
\ee
where $K(q, {\bar q})$ is an arbitrary function on $q^i$ and $q_j$.

As the last remark of this subsection, let us note that the ``tensor'' multiplet can be
constructed in terms of the ``hypermultiplet''.
Indeed, let us represent
$V^{ij}$  as the following composite superfield:
\be\label{tensor15}
\tV^{11}= -i\sqrt{2}\,q^1{\bar q}^1\; , \quad \tV^{22}
= -i\sqrt{2}\,q^2{\bar q}^2\;, \quad
\tV^{12}= -\frac{i}{\sqrt{2}}\left( q^1{\bar q}^2+ q^2{\bar q}^1 \right)\;.
\ee
One can check that, as a consequence of the ``hypermultiplet'' constraints \p{hyper4a},
the composite
superfield $\tV^{ij}$ automatically obeys \p{tensor4}. This is just the relation
established
in \cite{Olaf}.

The expressions \p{tensor15} supply a rather special solution
to the ``tensor'' multiplet constraints.
In particular, they express the auxiliary field of $\tV^{ij}$ through
the time derivative of
the physical components of $q^i$, which contains
no auxiliary fields.\footnote{This is a nonlinear version of
the phenomenon which holds in general for $d=1$ supersymmetry and
was discovered at the linearized level in \cite{GR,pato}.}
As a consequence, the superpotential of $\tV^{ik}$ is a particular
case of the $q^i$ superpotential,
which produces no genuine scalar potential for physical bosons and
gives rise for them only
to a Wess-Zumino type term
of the first order in the time derivative.

\subsubsection{N=4, d=1 chiral multiplet -- ${\bf (2,4,2)}$.}
The chiral $N=4$ supermultiplet is the simplest one. It can be described,
in full analogy with the $N=2$ case, by a complex superfield $\phi$ subjected
to the constraints \cite{{AP},{FR}}
\be\label{chir}
D^i \phi =0,\quad \bD_j {\bar \phi} =0.
\ee
The sigma-model type action for this multiplet
\be\label{actionchir}
S_\sigma = \int d^4\theta K(\phi,\bar\phi)
\ee
may be immediately extended to include the potential terms
\be\label{pot1}
S_{pot} = \int d^2\bar\theta F(\phi) + \int d^2\theta {\bar F}( \bar\phi).
\ee
In more details, such a supermultiplet
and the corresponding actions have been considered in \cite{leva}.

\subsubsection{The "old tensor" multiplet -- ${\bf (1,4,3)}$.}
The last possibility corresponds to the single bosonic superfield $u$.
In this case no inear constraints appear, since four fermionic components
are expressed through four spinor derivatives of $u$. As it was shown in \cite{leva},
one should impose some additional, second order in spinor derivatives, irreducibility constraints on $u$
\be\label{last}
D^i D_i\; e^{ u}= \bD{}_i\bD^i\; e^{ u}=
\left[ D^i,\bD_i \right] e^{ u} =0,
\ee
in order to pick up in $u$ the minimal off-shell field
content ({\bf 1},{\bf 4},{\bf 3}).
Once again, the detailed discussion of this case
can be found in \cite{leva}.

In fact,  we could re-derive the multiplet $u$ from the tensor multiplet $V^{ik}$ discussed in
the previous subsection. Indeed, one can construct the composite superfield
\be
e^{\tilde{u}} = \frac{1}{\sqrt{V^2}}\,, \label{last1}
\ee
which obeys just the constraints \p{last} as a consequence of \p{tensor4}.
The relation \p{last1} is
of the same type as the previously explored substitution
\p{tensor15}  and expresses two out
of the three auxiliary fields of $\tilde u$
via physical bosonic fields of $V^{ik}$ and time derivatives
thereof. The superconformal invariant action
for the superfield $u$
can be also constructed \cite{leva}
\be
S_{Conf}=\int dt d^4\theta \, e^{u}\, {u}\,.
\ee

By this, we complete the superfield description of all linear $N=4, d=1$ supermultiplets.
But the story about $N=4, d=1$ supermultiplets is not finished. There are nonlinear
supermultiplets which make everything much more interesting. Let us discuss here only one example -
the nonlinear chiral multiplet \cite{{ikl1},{bbkno}}.

\subsubsection{N=4, d=1 nonlinear chiral multiplet -- ${\bf (2,4,2)}$.}

The idea of nonlinear chiral supermultiplets comes about as follows.
If the two bosonic superfields $\cZ$ and $\cbZ$ parameterize the
two dimensional sphere $SU(2)/U(1)$ instead of flat space, then
they transform under $SU(2)/U(1)$ generators with the parameters
$a, \bar a$ as
\be\label{eq1a} \delta \cZ = a +{\bar a} \cZ{}^2
,\quad \delta \cbZ = {\bar a} +a \cbZ{}^2,
\ee
With respect to the
same group $SU(2)$, the $N=4$ covariant derivatives could form a
doublet
\be\label{eq2a} \delta D_i = - a \bD_i ,\quad \delta
\bD_i={\bar a} D_i \;.
\ee
One may immediately check that the
ordinary chirality conditions
$$ D_i \cZ =0 ,\quad \bD_i \cbZ=0 $$
are not invariant with respect to \p{eq1a}, \p{eq2a} and they should
be replaced, if we wish to keep the $SU(2)$ symmetry. It is rather
easy to guess the proper $SU(2)$ invariant constraints
\be\label{eq4a} D_i \cZ = - \alpha\cZ \bD_i \cZ ,\quad \bD_i \cbZ =
\alpha\cbZ D_i \cbZ \;,\qquad \alpha={\rm const}.
\ee
So, using the
constraints \p{eq4a}, we restore the $SU(2)$
invariance, but the price for this is just the nonlinearity of the
constraints. Let us stress that $N=4, d=1$ supersymmetry is
the minimal one where the constraints \p{eq4a} may appear,
because the covariant derivatives (and the supercharges) form a
doublet of $SU(2)$ which cannot be real.

The  $N=4, d=1$ nonlinear chiral supermultiplet involves one complex
scalar bosonic superfield $\cZ$ obeying the constraints
(\ref{eq4a}). If the real parameter $\alpha \neq 0$, it is always
possible to pass to $\alpha=1$ by a redefinition of the
superfields $\cZ, \cbZ$. So, one has only two essential values
$\alpha=1$ and $\alpha=0$. The latter case corresponds to the standard
$N=4, d=1$ chiral supermultiplet.
Now one can write the most general $N=4$ supersymmetric Lagrangian in $N=4$ superspace
\be\label{action1a}
S = \int\! dt d^2\theta d^2 \bar\theta\; K(\cZ,\cbZ)
+ \int\! dt d^2 \bar\theta\; F(\cZ) + \int\! dt
    d^2\theta\; \bF (\cbZ) \;.
\ee
Here $K(\cZ,\cbZ)$ is an arbitrary function of the superfields $\cZ$ and $\cbZ$, while $F (\cZ)$ and
$\bF(\cbZ)$ are arbitrary holomorphic functions depending only on $\cZ$ and $\cbZ$, respectively.
 Let us
stress that our superfields $\cZ$ and $\cbZ$ obey the nonlinear
variant of chirality conditions \p{eq4a}, but nevertheless the last
two terms in the action $S$ \p{action1a} are still invariant with
respect to the full $N=4$ supersymmetry. Indeed, the supersymmetry
transformations of the integrand of, for example, the second term in
\p{action1a} read
\be\label{dok1a} \delta F(\cZ) = -\epsilon^i D_i
F(\cZ)+2i\epsilon^i \bar\theta_i\dot{F}(\cZ)-\bar\epsilon_i
{\overline D}{}^i F(\cZ) +2i\bar\epsilon_i\theta^i\dot{F}(\cZ) \;.
\ee
Using the constraints \p{eq4a} the first term in the r.h.s. of
\p{dok1a} may be rewritten as
\be\label{dok2a} -\epsilon^i D_i F
=-\epsilon^i F_{\cZ} D_i \cZ= \alpha \epsilon^i F_{\cZ} \cZ \bD_i
\cZ \equiv \alpha \epsilon^i \bD_i \int d\cZ \; F_{\cZ} \cZ \;.
\ee
Thus, all terms in \p{dok1a} either are full time derivatives or
disappear after integration over $d^2 \bar\theta$.

The irreducible component content of $\cZ$, implied by \p{eq4a}, does not depend on $\alpha$ and can
be defined as
\be\label{compa}
    z = \cZ| , \; \bz = \cbZ|, \; A = -i \bD{}^i \bD_i \cZ| , \;
    \bar{A} =-i D^i D_i \cbZ| , \; \psi^i = \bD{}^i \cZ| , \; \bpsi{}^i = -D^i \cbZ|,
\ee where $|$ means restricting expressions to $\theta_i=\bar\theta{}^j=0$.
All higher-dimensional components are expressed as time
derivatives of the irreducible ones. Thus, the $N=4$ superfield
$\cZ$ constrained by \p{eq4a} has the same field content as the
linear chiral supermultiplet.

After integrating in \p{action1a} over the Grassmann variables and eliminating the auxiliary fields $A, \bar{A}$ by
their equations of motion, we get the action in terms of physical components
\be\label{action2a}
S = \int dt \left\{ g
    \dot{z} \dot{\bz} - i\alpha\frac{ \dot{z} \bz
    }{\fc}F_{z} + i\alpha\frac{\dot{\bz} z
    }{\fc}\bF_{\bz} - \frac{ F_{z} \bF_{\bz} }
    {g\fac^2} + \mbox{fermions} \right\},
\ee
where
\be\label{def1a}
 g(z,\bz)=\frac{\partial^2 K(z,\bz)}{\partial z\partial\bz},\quad
F_z=\frac{dF(z)}{dz},\quad \bF_\bz=\frac{d\bF (\bz )}{d\bz} \;.
\ee
{}From the  bosonic part of the action (\ref{action2a})
one may conclude that  the system contains a {\it nonzero magnetic field} with  the potential
\be
{\cal A}_{0}=i\alpha\frac{F_z \bz dz}{\fc} - i\alpha\frac{\bF_{\bz}zd{\bz}
    }{\fc}\;\;,\;
    d{\cal A}_{0}=i\alpha\frac{F_z+\bF_{\bz} }{(\fc)^2}dz\wedge d\bz.
\label{A0}
\ee
The strength of this magnetic field is given by the
expression
\be
B=\alpha\frac{(F_z+\bF_{\bz})}{(\fc)^2 g}\;.
\ee
As for the fermionic part of  the kinetic term, it can be represented as follows:
\be
 {\cal S}_{KinF}=\frac{i}{4}\int dt
(\fc )g\left(\psi\frac{D\bpsi}{dt}-\bpsi\frac{D\psi}{dt}\right),
\label{action3a}
\ee
where
\be
D\psi=d\psi+\Gamma\psi dz+T^+\bpsi dz,\quad D\bpsi=d\bpsi+{\bar\Gamma}\bpsi d\bz +T^-\psi d\bz,
\ee
and
\be\label{tor}
\Gamma=\partial_z\log \left((\fc){g}\right), \quad T^\pm=\pm\frac{\alpha}{\fc}.
\ee
Clearly enough, $\Gamma$, $\bar\Gamma$, $T^\pm$ define the components of the connection defining the configuration
superspace. The components  $\Gamma$ and $\bar\Gamma$ could be identified with the components
of the symmetric connection on the base space equipped with the metric $(\fc )gdzd\bz$, while
the rest does not have a similar interpretation.

Thus, we conclude that the main differences between the $N=4$ supersymmetric mechanics with nonlinear
chiral supermultiplet and
the standard one are the coupling of the fermionic degrees of freedom to the background, via the deformed connection,
the possibility to introduce a magnetic field, and the deformation of the
bosonic potential.

So far, we presented the results without explanations about how they were found. It appears to be desirable
to put the construction and study of $N$-extended supersymmetric models on a systematic basis by working
out the appropriate superfield techniques. Such a framework exists and is based on a superfield nonlinear
realization of the $d=1$ superconformal group. It was pioneered in \cite{leva} and recently advanced in \cite{ikl2,ikl1,bikl2}.
Its basic merits are,
firstly, that in most cases it automatically yields the irreducibility conditions for $d{=}1$ superfields
and, secondly, that it directly specifies the superconformal
transformation properties of these superfields. The physical bosons and fermions, together with the
$d{=}1$ superspace coordinates,  prove to be coset parameters associated with the appropriate generators of
the superconformal group. Thus, the differences in the field content of the various supermultiplets are
attributed to different choices of the coset supermanifold inside the given superconformal group.

\section{Nonlinear realizations}\label{sec:2}
In the previous section we considered the $N=4, d=1$ liner supermultiplets and constructed
some actions. But the most important questions concerning the irreducibility constraints
and the transformation properties of the superfields were given as an input. In this section we are
going to demonstrate that most of the constraints and all transformation properties can be
obtained automatically as results of using the nonlinear realization approach. Our consideration
will be mostly illustrative -- we will skip presenting any proofs. Instead, we will pay attention
to the ideas and technical features of this approach.

\subsection{Realizations in the coset space}
The key statement of the nonlinear realization approach may be formulated as follows:
\begin{theorem}
If a group $G$ acts transitively\footnote{This means that the transformations from $G$ relate any two
arbitrary points in the space.} on some space, and the subgroup $H$ preserves a given point of this
space, then this action of the group $G$ may be realized by left multiplications on the coset $G/H$,
while the coordinates which parameterize  the coset $G/H$ are just the coordinates of the space.
\end{theorem}

As the simplest example let us consider the four-dimensional Poincar\'e group $\left\{ P_\mu, M_{\mu\nu}\right\}$.
Is is clear that the transformations which preserve some point are just rotations around this points.
In other words, $H=\left\{ M_{\mu\nu}\right\}$. Therefore, due to our Theorem, a natural realization
of the Poincar\'e group can be achieved in the coset $G/H$. This coset contains only the translations $P_\mu$,
and it is natural to parameterized this coset as
\be\label{nl1}
G/H= e^{ix^\mu P_\mu}.
\ee
It is evident that the Poincar\'e group may be realized on the four coordinates $\left\{ x^\mu\right\}$. What is
important here is that we do not need to know how our coordinates transform under the Poincar\'e group. Instead,
we can deduce these transformations from the representation \p{nl1}.

Among different cosets there are special ones which are called orthonormal. They may be described as follows.
Let us consider the group $G$ with generators $\left\{ X_i, Y_\alpha \right\}$ which obey the following
relations:
\bea\label{nl2}
&&\left[ Y_\alpha, Y_\beta \right] =i C^\gamma_{\alpha\beta} Y_\gamma, \nn
&&\left[ Y_\alpha, X_i \right] =i C^j_{\alpha i } X_j + i C^\beta_{\alpha i} Y_\beta, \nn
&&\left[ X_i, X_j \right] =i C^k_{ij } X_k + i C^\alpha_{ij} Y_\alpha,
\eea
where $C$ are structure constants. We see that the generators $Y_\alpha$ form the subgroup $H$. The coset
$G/H$ is called orthonormal if $C^\beta_{\alpha i}=0$. In other words, this property means that the
generators $X_i$ transform under some representation of the stability subgroup $H$. In what follows we
will consider only such a coset. A more restrictive class of cosets -- the symmetric spaces -- corresponds
to the additional constraints $C^k_{ij }=0$.

A detailed consideration of the cosets and their geometric properties may be found in \cite{nl}.

A very important class is made by the cosets which contain space-time symmetry generators as well as
the generators of internal symmetries. In order to deal with such cosets we must
\begin{itemize}
\item introduce the coordinates for space-time translations (or/and super-translations);
\item introduce the parameters for the rest of the generators in the coset. These additional
parameters are treated as {\it fields} which depend on the space-time coordinates.
\end{itemize}
The fields (superfields) which appear as parameters of the coset will have inhomogeneous transformation
properties. They are known as Goldstone fields. Their appearance is very important: they are definitely
needed to construct an action, which is invariant with respect to transformations from the coset $G/H$.
Let us repeat this point: in the nonlinear realization approach only the $H$-symmetry is manifest. The invariance
under $G/H$ transformations  is achieved through the interaction of matter fields with Goldstone ones.

Finally, let us note that the number of essential Goldstone fields does not always coincide with the number
of coset generators. As we will see later, some of the Goldstone fields often can be expressed through
other Goldstone fields. This is the so called Inverse Higgs phenomenon \cite{InvH}.

Now it is time to demonstrate how all this works on the simplest examples.

\subsection{Realizations: Examples \& Technique}
\subsubsection{N=2, d=1 Super Poincar\'e.}
Let us start with the simplest example of the $N=2, d=1$ super Poincar\'e algebra which contains
two super-translations $Q, \bar{Q}$ which anticommute on the time-translation $P$ \p{22}
$$ \left\{ Q, \bar{Q} \right\} =-2P.$$
In this case the stability subgroup is trivial and all generators are in the coset. Therefore,
one should introduce coordinates for all generators:
\be\label{nl3}
g=G/H=e^{itP} e^{\theta Q +\bar\theta {\bar Q}} \;.
\ee
These coordinates $\left\{ t, \theta,\bar\theta \right\}$ span $N=2, d=1$ superspace. Now we are going to
find the realization of $N=2$ superalgebra \p{22} in this superspace. The first step is to find the
realization of the translation $P$. So, we act on the element \p{nl3} from the left by $g_0$
\be\label{nl4}
g_0=e^{iaP}: \; g_0\cdot g = e^{iaP}\cdot e^{itP} e^{\theta Q +\bar\theta {\bar Q}} =
e^{i(t+a)P} e^{\theta Q +\bar\theta {\bar Q}} \equiv e^{it'P} e^{\theta' Q +\bar\theta{}' {\bar Q}}.
\ee
Thus, we get the standard transformations of the coordinates
\be\label{nl5}
P: \quad \left\{ \begin{array}{l}
\delta t= a \\
\delta \theta= \delta\bar\theta =0. \end{array} \right.
\ee
Something more interesting happens for the supertranslations
\be\label{nl6}
g_1=e^{\epsilon Q+ \bar\epsilon \bar{Q}}: \; g_1\cdot g =
e^{\epsilon Q+ \bar\epsilon \bar{Q}}\cdot e^{itP} e^{\theta Q +\bar\theta {\bar Q}} =
e^{itP}e^{\epsilon Q+ \bar\epsilon \bar{Q}} e^{\theta Q +\bar\theta {\bar Q}} .
\ee
Now we need to bring the product of the exponents in \p{nl6} to the standard form
$$e^{it'P} e^{\theta' Q +\bar\theta{}' {\bar Q}}.$$
In order to do this, one should use the Campbell-Hausdorff formulae
\be\label{CH}
e^A \cdot e^B = \exp \left( A+B+\frac{1}{2} \left[ A,B\right] +\frac{1}{12}\left[ A,\left[ A,B\right]\right]-
\frac{1}{12}\left[ \left[ A,B\right],B\right]+\ldots \right)
\ee
Thus we will get
\be\label{nl7}
e^{\epsilon Q+ \bar\epsilon \bar{Q}} e^{\theta Q +\bar\theta {\bar Q}} =e^{(\theta+\epsilon) Q+
(\bar\theta+\bar\epsilon) \bar{Q} + (\epsilon\bar\theta+\bar\epsilon \theta)P}.
\ee
So, the supertranslation is realized as follows:
\be\label{nl8}
Q, \bar{Q} : \quad \left\{ \begin{array}{l}
\delta t= -i (\epsilon\bar\theta+\bar\epsilon \theta) \\
\delta \theta=\epsilon, \; \delta\bar\theta =\bar\epsilon. \end{array} \right.
\ee
This is just the transformation \p{32} we used before. Of course, in this rather simple case we could
find the answer without problems. But the lesson is that the same procedure works always, for
any (super)group and any coset. Let us consider a more involved example.

\subsubsection{d=1 Conformal group.}
The conformal algebra in $d=1$ contains three generators: translation $P$, dilatation $D$ and
conformal boost $K$ obeying to the following relations:
\be\label{nl9}
i\left[ P,K \right] =-2D, \; i\left[ D,P \right] =P, \; i\left[ D,K \right] =-2K.
\ee
The stability subgroup is again trivial and our coset may be parameterized as
\be\label{nl10}
g=e^{itP} e^{i z(t) K} e^{i u(t) D}\;.
\ee
Let us stress that we want to obtain a one dimensional realization. Therefore, we can
introduce only one coordinate - time $t$. But we have three generators in the coset.
The unique solution\footnote{Of course, we may put some of the generators $K$ and $D$,
or even both of them, in the stability subgroup $H$. But in this case all matter fields
should realize a representation of $H=\left\{K,D\right\}$ which never happens.}
is to consider the two other coordinates as functions of time.
Thus, we have to introduce two fields $z(t)$ and $u(t)$ which are just Goldstone fields.

Let us find the realization of the conformal group in our coset \p{nl10}.

The translation is realized trivially, as in \p{nl5}, so we will start from the dilatation
\be\label{nl11}
g_0=e^{iaD}: \; g_0\cdot g = e^{iaD}\cdot e^{itP} e^{i z(t) K} e^{i u(t) D} .
\ee
Now we have a problem - how to commute the first exponent in \p{nl11} with the remaining ones?
The Campbell-Hausdorff formulae \p{CH} do not help us too much, because the series does not terminate.
A useful trick is to represent r.h.s. of \p{nl11} in the form
\be\label{nl12}
\underbrace{e^{iaD} e^{itP} e^{-iaD}}_1 \underbrace{e^{iaD} e^{i z(t) K}e^{-iaD}}_2 e^{iaD}  e^{i u(t) D}.
\ee
In order to evaluate \p{nl12} we will use the following formulas due to Bruno Zumino \cite{zumino}:
\be\label{BZ}
e^A B e^{-A} \equiv e^A \wedge B,\quad e^A e^B e^{-A} \equiv e^{e^A \wedge B},
\ee
where
\be
A^n \wedge B \equiv \underbrace{\left[A,\left[A,\left[\ldots\right.\right.\right.}_{n} \left.\left.\left.
,B\right]\right]\ldots\right]
\ee
Using \p{BZ} we can immediately find
\be\label{nl13}
D: \quad \left\{ \begin{array}{l}
\delta t= at \\
\delta u= a,\quad \delta z =-a z \end{array} \right.
\ee
We see that the field $u(t)$ is shifted by the constant parameter $a$ under dilatation. Such Goldstone
field is called {\it dilaton}.

Finally, we should find the transformations of the coordinates under conformal boost $K$:
\be\label{nl14}
g_1=e^{ibK}: \; g_1\cdot g = \underbrace{e^{ibK} e^{itP}}_1 e^{i z(t) K} e^{i u(t) D} .
\ee
Here, in order to commute the first two terms we will use the following trick:
let us represent the first two exponents in the r.h.s. of \p{nl14} as follows:
\be\label{nl15}
e^{ibK} e^{itP}=e^{itP}\cdot \tilde{g}\; \Rightarrow \; \tilde{g}=e^{-itP}e^{ibK} e^{itP}.
\ee
So, we again can use \p{BZ} to calculate\footnote{We are interested in the infinitesimal transformations
and omit all terms which are higher then linear in the parameters.} $\tilde{g}$
$$
\tilde{g}=e^{ibK+2ibtD+ibt^2P}\approx e^{ibK}e^{2ibtD}e^{ibt^2P} \;.
$$
Thus, we have
\be\label{nl16}
K: \quad \left\{ \begin{array}{l}
\delta t= bt^2 \\
\delta u= 2bt,\quad \delta z =b-2bt z \end{array} \right.
\ee

Now we know how to find the transformation properties of the coordinates and
Goldstone (super)fields for any cosets. The next important question is how to construct
the invariant and/or covariant objects.

\subsection{Cartan's forms}
The Cartan's forms for the coset $g=G/H$ are defined as follows:
\be\label{cf1}
g^{-1} dg = i \omega^i X_i +i \omega^\alpha Y_\alpha,
\ee
where the generators $\left\{ X_i, Y_\alpha \right\}$ obey to \p{nl2}.

By the definition \p{cf1} the Cartan's forms are invariant with respect to left
multiplication of the coset element $g$. Let us represent the result of the left
multiplication of the coset element $g$ as follows:
\be\label{cf2}
g_0 \cdot g = \tilde{g} \cdot h \;.
\ee
Here,  $\tilde{g}$ belongs to the coset, while the element $h$ lies in the stability
subgroup $H$. Now we have
\be\label{cf3}
i \omega^i X_i +i \omega^\alpha Y_\alpha =h^{-1}
\underbrace{\left( \tilde{g}{}^{-1}d \tilde{g}\right)}_{i \tilde\omega{}^i X_i +i \tilde\omega{}^\alpha Y_\alpha}h+h^{-1}dh .
\ee
If the coset $g=G/H$ is othonormal, then
\bea\label{cf4}
&& \tilde\omega{}^i X_i =h \cdot\omega{}^i X_i \cdot h^{-1} ,\nn
&& \tilde\omega{}^\alpha Y_\alpha = h \cdot\omega{}^\alpha Y_\alpha\cdot h^{-1}+i dh \cdot h^{-1}.
\eea
Thus we see that the forms $\omega{}^i$ which belong to the coset transform homogeneously,
while the forms $\omega{}^\alpha$ on the stability subgroup transform like connections and
can be used to construct covariant derivatives.

Finally, one should note that, in the exponential parametrization we are using through
these lectures, the evaluation of the Cartan's forms is based on the following identity \cite{zumino}:
\be\label{cf5}
e^{-A} d e^A = \frac{1-e^{-A}}{A} \wedge dA .
\ee
Now it is time for some examples.

\subsubsection{N=2, d=1 Super Poincar\'e}
Choosing the parametrization of the group element as in \p{nl3} one may immediately find the Cartan's forms
\be\label{cf6}
\omega_P = dt-i\left( d\bar\theta \theta+d\theta\bar\theta\right),\quad \omega_Q=d\theta,\; \bar{\omega}_Q=d\bar\theta.
\ee
Thus, the Cartan's forms \p{cf6} just coincide with the covariant differentials \p{36} we guessed before.

\subsubsection{d=1 Conformal group}
The Cartan's forms for the coset \p{nl10} may be easily calculated using \p{cf5}
\be\label{cf7}
\omega_P=e^{-u},\; \omega_D=du-2 z dt,\; \omega_K =e^u \left[ dz + z^2 dt\right].
\ee
What is the most interesting here is the structure of the form $\omega_D$. Indeed, from the previous
consideration we know that $\omega_D$ being coset forms, transforms homogeneously. Therefore, the
following condition:
\be\label{cf8}
\omega_D=0 \; \Rightarrow \; z= \frac{1}{2} \dot{u}
\ee
is invariant with respect to the whole conformal group! This means that the Goldstone field $z(t)$ is unessential
and can be expressed in terms of the dilaton $u(t)$. This is the simplest variant of the inverse Higgs phenomenon \cite{InvH}.
With the help of the Cartan' forms \p{cf8} one could construct the simplest invariant action
\bea\label{cf9}
S&=&-\int \left( \omega_K + m^2 \omega_P\right) = \int dt \left[ \frac{1}{4} e^u \dot{u}{}^2 - m^2 e^{-u}\right] = \nn
  &&  \int dt \left[ \dot{\rho}{}^2 - \frac{m^2}{\rho^2}\right], \quad \rho \equiv e^{\frac{u}{2}},
\eea
with $m$ being a parameter of the dimension of mass.
The action \p{cf9} is just the  conformal mechanics action \cite{dff}.

It is rather interesting that we could go a little bit further.

The basis \p{nl9} in the conformal algebra $so(1,2)$ can naturally be called ``conformal'',
as it implies the standard $d=1$ conformal transformations for the time $t$.
Now we pass to another basis in the same algebra
\be\label{adsgenerators}
\hK =mK -\frac{1}{m}P\;,\; \hD=mD \;.
\ee
This choice will be referred to as
the ``AdS basis'' for a reason soon to be made clear soon.

The conformal algebra \p{nl9} in the AdS basis \p{adsgenerators}
reads
\be\label{adsbasis}
i \left[ P,\hD\right] =-mP\; , \; i \left[ \hK,\hD\right] =2P+ m\hK\; , \;i \left[ P,\hK
 \right] =-2\hD\; .
\ee
An element of $SO(1,2)$ in the AdS basis is defined to be
\be\label{confcoset}
g=e^{iyP}e^{i\phi(y) \hD} e^{i\Omega(y) \hK} \; .
\ee

Now we are in a position to explain the motivation for the nomenclature ``AdS basis''.
The generator
$\hK$ \p{adsgenerators} can be shown to correspond to a $SO(1,1)$
subgroup of $SO(1,2)$. Thus the parameters $y$ and $\phi(y)$ in \p{confcoset}
parameterize the coset $SO(1,2)/SO(1,1)$, i.e. AdS$_2$. The parametrization
\p{confcoset} of AdS$_2$ is a particular case of the so-called ``solvable
subgroup parametrization'' of the AdS spaces. The $d=4$ analog
of this parametrization is the parametrization of the AdS$_5$ space
in such a way that its coordinates are still parameters associated with
the 4-translation and dilatation generators $P_m, D$ of $SO(2,4)$, while it
is the subgroup $SO(1,4)$ with the algebra $ \propto \{P_m - K_m, so(1,3)\}$
which is chosen as the stability subgroup .

The difference in the geometric meanings of the coordinate pairs $\left(t,
u(t)\right)$ and $\left(y, \phi(y)\right)$ is manifested in their different transformation
properties under the same $d=1$ conformal transformations. Left shifts of
the $SO(1,2)$ group element in the parametrization \p{confcoset} induce the
following transformations:
\be
\delta y = a(y) + \frac{1}{m{}^2}\,c\,e^{2m\phi}~, \quad \delta \phi = \frac{1}{m}\,
\dot{a}(y) = \frac{1}{m}\,(b+2c\,y)~, \quad \delta \Omega = \frac{1}{m}\,c\,e^{m\phi} ~.
\label{modphitr}
\ee
We observe the modification of the special conformal transformation of $y$ by
a field-dependent term.

The relevant left-invariant Cartan forms are given by the following expressions:
\bea\label{confcartan}
&& \hat\omega_D= \frac{1+\Lambda^2}{1-\Lambda^2} d\phi -2 \frac{\Lambda}{1-\Lambda^2}
 e^{-m\phi}dy~, \nn
&& \hat\omega_P =\frac{1+\Lambda^2}{1-\Lambda^2} e^{-m\phi} d y - 2
 \frac{\Lambda}{1-\Lambda^2} d\phi~, \nn
&& \hat\omega_K = m\frac{\Lambda}{1-\Lambda^2}\left( \Lambda e^{-m\phi} dy - d\phi\right) +
\frac{d\Lambda}{1-\Lambda^2}\,~,
\eea
where
\be\label{deflabda}
\Lambda = \tanh \Omega \;.
\ee

As in the previous realization, the field $\Lambda(y)$ can be eliminated by imposing the
inverse Higgs constraint
\be\label{confih}
\hat\omega_D=0 \; \Rightarrow \; \partial_y \phi = 2\,e^{-m\phi}\,
 \frac{\Lambda}{1+\Lambda^2} \;,
\ee
whence $\Lambda $ is expressed in terms of $\phi $
\be
\Lambda = \partial_y\phi\,e^{m\phi}\,\frac{1}{1 + \sqrt{1 - e^{2m\phi}\,
 (\partial_y\phi)^2}}~. \label{Lamb}
\ee

The $SO(1,2)$ invariant distance on AdS$_2$ can be defined, prior to imposing any
constraints, as
\be
ds^2 = -\hat\omega_P^2 + \hat\omega_D^2 = -e^{-2m\phi}\,dy^2 + d\phi^2~.
\label{dist1}
\ee
Making the redefinition
$$
U = e^{-m\phi}~,
$$
it can be cast into the standard Bertotti-Robinson metrics form
\be
ds^2 = -U^2\, dy^2 + (1/m^2)U^{-2}dU^2~, \label{dist2}
\ee
with $1/m$ as the inverse AdS$_2$ radius,
\be
{1\over m} = R~.
\ee

The invariant action can now be constructed from the new Cartan forms \p{confcartan}
which, after substituting the inverse Higgs expression for $\Lambda $, eq. \p{Lamb},
read
\bea
&& \hat\omega_P = e^{-m\phi}\,\sqrt{1 - e^{2m\phi}\,(\partial_y\phi)^2 }\;dy~, \nn
&& \hat\omega_K = -\frac{m}{2}\,e^{-m\phi}\left(1 - \sqrt{1 - e^{2m\phi}\,(\partial_y\phi)^2
}\right)dy
+ \mbox{Tot. deriv.}\times dy~. \label{invHforms}
\eea
The invariant action reads
\be
S = -\int \left(\tilde\mu \,\hat\omega_P - qe^{-m\phi} \right) =
-\int dy\, e^{-m\phi}\left(\tilde\mu\, \sqrt{1 -  e^{2m\phi}\,(\partial_y\phi)^2}
- q\right)~. \label{invact2}
\ee
After the above field redefinitions it is recognized as the radial-motion part of the
``new'' conformal mechanics action \cite{nscm}. Notice that the second term in \p{invact2}
is invariant under \p{modphitr}, up to a total derivative in the integrand. The
action can be rewritten in a manifestly invariant form (with a tensor Lagrangian) by using
the explicit expression for $\hat\omega_K$ in \p{invHforms}
\be
S = \int \left[ (q-\tilde\mu)\,\hat\omega_P - (2/m)q\,\hat\omega_K \right]~.
\label{stens}
\ee

Now we are approaching the major point. We see that the ``old'' and ``new''
conformal mechanics models are associated with two different nonlinear
realizations of the same $d=1$ conformal group $SO(1,2)$ corresponding, respectively,
to the two different  choices \p{nl9} and \p{confcoset} of the parametrization of the
group element.
The invariant actions in both cases can be written as integrals of linear combinations
of the left-invariant Cartan forms. But the latter {\it cannot} depend on the
choice of parametrization. Then the actions \p{cf9} and \p{invact2}
should in fact coincide with each other, up to a redefinition of the free parameters of the
actions. Thus
two conformal mechanics models are {\it equivalent} modulo redefinitions of the
involved time coordinate and field. This statement should be contrasted with the
previous view of the ``old'' conformal mechanics model as a ``non-relativistic''
approximation of the ``new'' one.

\subsection{Nonlinear realizations and supersymmetry}
One of the interesting applications of the nonlinear realizations technique is
that of establishing irreducibility constraints for superfields.

In the present section we focus on the case of $N{=}4, d{=}1$ supersymmetry
(with 4 real supercharges) and propose to derive its various irreducible off-shell
superfields from different
nonlinear realizations of the most general $N{=}4, d{=}1$ superconformal
group $D(2,1;\alpha)$.
An advantage of this approach is that it simultaneously specifies the
superconformal transformation
properties of the superfields, though the latter can equally be used for
constructing non-conformal
supersymmetric models as well. As the essence of these techniques, any given
irreducible $N{=}4, d{=}1$ superfield
comes out
as a Goldstone superfield parameterizing, together with the
$N{=}4, d{=}1$ superspace coordinates,
some supercoset of $D(2,1;\alpha)$. The method was
already employed in the paper \cite{Olaf}
where the off-shell multiplet $({\bf 3, 4, 1})$
was re-derived from the nonlinear realization
of  $D(2,1;\alpha)$ in the coset with an
$SL(2,R)\times \left[SU(2)/U(1)\right]$ bosonic part
(the second $SU(2)\subset D(2,1; \alpha)$ was placed into the stability subgroup).

Here we consider nonlinear realizations of the same conformal
supergroup  $D(2,1;\alpha)$
in its other coset superspaces. In this way we reproduce
the $({\bf 4, 4, 0})$ multiplet
and also derive two new nonlinear off-shell multiplets.
The $({\bf 4,4,0})$ multiplet is represented by superfields
parameterizing a supercoset with the bosonic part being $SL(2,R)\times SU(2)$,
where the dilaton and the three parameters
of $SU(2)$ are identified with the four physical bosonic fields.
One of the new Goldstone multiplets is a $d{=}1$ analog of the so-called
nonlinear multiplet of $N{=}2, d{=}4$ supersymmetry.
It has the same off-shell contents $({\bf 3,4, 1})$ as the multiplet
employed in \cite{Olaf},
but it obeys a different constraint and enjoys different superconformal
transformation properties. It corresponds to the specific nonlinear
realization of
$D(2,1; \alpha)$, where the dilatation generator and one of the two $SU(2)$
subgroups are placed
into the stability subgroup. One more new multiplet of a similar type
is obtained by placing into the stability subgroup,
along with the dilatation and three $SU(2)$ generators,
also the $U(1)$
generator from the second $SU(2) \subset D(2,1;\alpha)$. It has the same
field content as
a chiral $N{=}4, d{=}1$ multiplet, i.e. $({\bf 2, 4, 2})$. Hence, it may be termed
as the nonlinear chiral
supermultiplet. It is exceptional, in the sense that no analogs for it are known
in $N{=}2, d{=}4$
superspace.

\subsubsection{Supergroup D(2,1;$\alpha$) and its  nonlinear realizations}
We use the standard definition  of the superalgebra $D(2,1;\alpha)$
with the notations of ref. \cite{Olaf}.
It contains nine bosonic generators which form a direct sum of $sl(2)$
with generators
$P,D,K$ and two $su(2)$ subalgebras with generators
$V, \bV, V_3\, \; \mbox{ and }
\; T, \bT, T_3$, respectively
\bea\label{alg1}
&& i\left[ D,P\right] =P,\;  i\left[ D,K\right]=-K ,\;
i\left[ P,K\right]=-2D , \quad
i\left[ V_3,V\right]=-V,\;  i\left[ V_3,\bV \right]=\bV, \nn
&& i\left[ V,\bV\right]=2V_3,\quad
i\left[ T_3,T\right]=-T,\;  i\left[ T_3,\bT \right]=\bT,\;
i\left[ T,\bT\right]=2T_3.
\eea
The eight fermionic generators $Q^i,\bQ_i,S^i,\bS_i$
are in the fundamental representations of all bosonic subalgebras
(in our notation only
one $su(2)$ is manifest, viz. the one with generators
$V, \bV, V_3$)
\bea\label{alg2}
&&i\left[D ,Q^i \right] = \frac{1}{2}Q^i,\;
i\left[D ,S^i \right] = -\frac{1}{2}S^i, \quad
i\left[P ,S^i \right] =-Q^i,\;
i\left[K ,Q^i \right] =S^i, \nonumber \\
&& i\left[V_3 ,Q^1 \right] =\frac{1}{2}Q^1,\; i\left[V_3 ,Q^2 \right]
=-\frac{1}{2}Q^2,\quad
i\left[V ,Q^1 \right] =Q^2, \; i\left[V ,\bQ_2 \right] =-\bQ_1, \nonumber \\
&&i\left[V_3 ,S^1 \right] =\frac{1}{2}S^1,\;    i\left[V_3 ,S^2 \right]
=-\frac{1}{2}S^2, \quad
i\left[V ,S^1 \right] =S^2, \;    i\left[V ,\bS_2 \right] =-\bS_1,  \nonumber\\
&& i\left[T_3 ,Q^i\right] =\frac{1}{2}Q^i, \;
i\left[T_3 ,S^i\right] =\frac{1}{2}S^i, \quad
i\left[T ,Q^i\right] =\bQ^i, \;
i\left[T ,S^i\right] =\bS^i
\eea
(and c.c.). The splitting of the fermionic
generators into the $Q$ and $S$ sets is natural and useful, because
$Q^i,\bQ_k$ together with
$P$ form $N=4, d=1$ super Poincar\'e subalgebra, while $S^i,\bS_k $ generate
superconformal translations
\be\label{allg3}
\left\{Q^i ,\bQ_j \right\} = -2\delta^i_j P , \quad \left\{S^i ,\bS_j \right\}
=-2\delta^i_j K .
\ee
The non-trivial dependence of the superalgebra $D(2,1;\alpha)$ on the parameter
$\alpha$
manifests itself only in the cross-anticommutators of the Poincar\'e and conformal
supercharges
\bea\label{alg4}
&& \left\{ Q^i,S^j \right\} =-2(1+\alpha )\epsilon^{ij} \bT , \;
\left\{Q^1 ,\bS_2 \right\} =2\alpha \bV ,\;\left\{Q^1 ,\bS_1 \right\}
=-2D-2\alpha V_3+2(1+\alpha)T_3 ,
               \nonumber \\
&& \left\{Q^2 ,\bS_1 \right\} =-2\alpha V, \;\left\{Q^2 ,\bS_2 \right\}
=-2D +2\alpha V_3+2(1+\alpha)T_3
\eea
(and c.c.). The generators $P,D,K$ are chosen to be hermitian, and the remaining ones obey
the following
conjugation rules:
\be\label{conjug}
\left( T \right)^\dagger = \bT, \; \left( T_3\right)^\dagger =-T_3 , \;
\left( V \right)^\dagger = \bV, \; \left( V_3\right)^\dagger =-V_3 , \;
\overline{\left( Q^i \right)}=\bQ_i,\; \overline{\left( S^i \right)}=\bS_i.
\ee

The parameter $\alpha $ is an arbitrary real number. For $\alpha = 0$ and
$\alpha = -1$ one of the
$su(2)$ algebras decouples and the superalgebra $su(1,1\vert 2)\oplus su(2)$
is recovered.
The superalgebra $D(2,1;1)$ is isomorphic
to $osp(4^*\vert 2)\,$.

We will be interested in diverse nonlinear realizations  of the
superconformal group $D(2,1;\alpha)$
in its coset superspaces. As a starting point we shall consider the following
parametrization of the supercoset:
\be\label{coset}
g=e^{itP}e^{\theta_i Q^i+\bt^i \bQ_i}e^{\psi_i S^i+\bpsi^i \bS_i}
e^{izK}e^{iuD}e^{i\varphi V+ i\bar\varphi \bV}e^{\phi V_3}~.
\ee
The coordinates $t, \theta_i, \bt^i$ parameterize the $N=4, d=1$ superspace. All other
supercoset parameters are Goldstone $N=4$ superfields. The  group
$SU(2)\propto \left( V,\bV,V_3\right)$
linearly acts on the doublet indices $i$ of spinor coordinates and Goldstone
fermionic superfields,
while the bosonic Goldstone superfields $\varphi, \bar\varphi, \phi$ parameterize
this $SU(2)$.
Another $SU(2)$, as a whole, is placed in the stability subgroup and acts only on
fermionic Goldstone superfields and $\theta$'s, mixing them with their conjugates.
With our choice of the $SU(2)$ coset, we are led to assume that $\alpha \neq 0$.

The left-covariant Cartan one-form $\Omega$ with values in the superalgebra
$D(2,1;\alpha)$
is defined by the standard relation
\be \label{cformdef}
g^{-1}\,d\,g = \Omega~.
\ee
In what follows we shall need the explicit structure of several
important one-forms in the expansion of $\Omega$ over the generators,
\bea\label{cforms}
&& \omega_D = idu-2\left( \bpsi^i d\theta_i + \psi_i d\bt^i \right)
  -2iz d{\tilde t} \; , \nn
&& \omega_V=\frac{e^{-i\phi}}{ 1 +\Lambda\bLam}\left[ id\Lambda+{\hat \omega}_V+
 \Lambda^2{\hat\bomega}_{V}-\Lambda{\hat\omega}_{V_3}\right], \;
 \bomega_{V}=\frac{e^{i\phi}}{ 1{+}\Lambda\bLam}\left[ id\bLam+{\hat \bomega}_{V}+
 \bLam^2{\hat\omega}_{V}+\bLam{\hat\omega}_{V_3}\right],\nn
&& \omega_{V_3}=d\phi+\frac{1}{ 1{+}\Lambda\bLam}\left[ i\left( d\Lambda\bLam -
\Lambda d\bLam\right)
+\left( 1{-}\Lambda\bLam\right){\hat\omega}_{V_3}
   -2\left( \Lambda{\hat\bomega}_{V}-\bLam{\hat\omega}_V \right)\right].
\eea
Here
\bea\label{cforms1}
&& {\hat\omega}_V=2\alpha \left[ \psi_2 d\bt^1 -\bpsi^1\left( d\theta_2-
       \psi_2 d{\tilde t}\right)\right], \;
 {\hat\bomega}_{V}=2\alpha \left[ \bpsi^2 d\theta_1 -\psi_1
    \left( d\bt^2-\bpsi^2 d{\tilde t}\right)\right],\nn
&& {\hat\omega}_{V_3}=2\alpha\left[ \psi_1 d\bt^1 -\bpsi^1 d \theta_1-
   \psi_2 d\bt^2 +\bpsi^2 d \theta_2 +
 \left( \bpsi^1 \psi_1 -\bpsi^2 \psi_2\right)d{\tilde t}\right],
\eea
\be
d{\tilde t} \equiv dt +i\left( \theta_id\bt^i+\bt^id\theta_i\right) \;,
\ee
and
\be \label{Lambda}
\Lambda = \frac{ \tan \sqrt{\varphi\bar\varphi}}{\sqrt{\varphi\bar\varphi}}\varphi
\; ,\;
\bLam = \frac{ \tan \sqrt{\varphi\bar\varphi}}{\sqrt{\varphi\bar\varphi}}\bar\varphi
\; .
\ee
The semi-covariant (fully covariant only under Poincar\'e supersymmetry)
spinor derivatives are defined by
\be
D^i=\frac{\partial}{\partial\theta_i}+i\bt^i \partial_t\; , \;
\bD_i=\frac{\partial}{\partial\bt^i}+i\theta_i \partial_t\; , \;
\left\{ D^i, \bD_j\right\}= 2i \delta^i_j \partial_t \; .
\ee

Let us remind that the transformation properties of the $N=4$ superspace coordinates
and the basic Goldstone superfields under the transformations of the supergroup
$D(2,1;\alpha)$ could be easily found, as we did in the previous sections.
Here  we  give the explicit expressions only for the variations
of our superspace
coordinates and superfields, with respect to two $SU(2)$ subgroup. They are generated
by the left
action of the group element
\be
g_0=e^{iaV+i{\bar a}\bV} e^{ibT+i{\bar b}\bT}
\ee
and read
\bea\label{su2}
&&\delta\theta_1={\bar b}\bt^2-{\bar a}\theta_2,\quad \delta\theta_2
=-{\bar b}\bt^1+a\theta_1 ,\nonumber \\
&& \delta \Lambda=a+{\bar a}\Lambda^2,\; \delta\bLam={\bar a}+a\bLam^2,\quad
\delta\phi=i\left( a\bLam-{\bar a}\Lambda\right) .
\eea

\subsubsection{N=4, d=1 ``hypermultiplet''}
The basic idea of our method is to impose the appropriate $D(2,1;\alpha)$ covariant
constraints on
the Cartan forms \p{cformdef}, \p{cforms}, so as to end up with some minimal
$N=4, d=1$ superfield
set carrying
an irreducible off-shell multiplet of $N=4, d=1$ supersymmetry.
Due to the covariance of the
constraints, the ultimate Goldstone superfields will support the corresponding
nonlinear realization of
the superconformal group  $D(2,1;\alpha)$.

Let us elaborate on this in some detail. It was the desire to keep $N=4, d=1$
Poincar\'e supersymmetry
unbroken that led us to associate the Grassmann coordinates
$\theta_i, \bar\theta^i$ with the
Poincar\'e supercharges in \p{coset} and the fermionic Goldstone superfields
$\psi_i, \bar\psi^i$ with
the remaining four supercharges which generate conformal supersymmetry.
The minimal number of physical
fermions in an irreducible $N=4, d=1$ supermultiplet is four, and it nicely matches
with the number
of fermionic Goldstone superfields in \p{coset}, the first components of which
can so be naturally
identified with the fermionic fields of the ultimate Goldstone supermultiplet.
On the other hand, we can vary
the number of bosonic Goldstone superfields in \p{coset}: by putting some of them
equal to zero we can
enlarge the stability subgroup by the corresponding generators and so switch
to another coset with a
smaller set of parameters. Thus, for different choices of the stability
subalgebra, the coset \p{coset}
will contain a different number of bosonic superfields, but always the
same number of fermionic superfields
$\psi_i,\bar\psi^i$. Yet, the corresponding sets of bosonic and fermionic
Goldstone superfields contain too many field components, and it is natural
to impose on
them the appropriate covariant constraints, in order to reduce the number
of components, as much as possible.
For preserving off-shell $N=4$ supersymmetry these constraints must be purely
kinematical, i.e. they must
not imply any dynamical restriction, such as equations of motion.

Some of the constraints just mentioned above should express the Goldstone fermionic
superfields in terms
of spinor derivatives of the bosonic ones. On the other hand, as soon
as the first components of
the fermionic superfields $\psi_i,\bar\psi^k$
are required to be the only physical fermions, we are led to impose much the stronger
condition
that {\it all} spinor derivatives of {\it all} bosonic superfields
be properly expressed in terms of
$\psi_i,\,\bar\psi^i\,$. Remarkably, the latter conditions
will prove to be
just the irreducibility constraints picking up irreducible $N=4$
supermultiplets.

Here we will consider in details only the most general case when the coset
\p{coset} contains all four bosonic
superfields $u,\varphi,\bar\varphi,\phi$. Looking at the structure of
the Cartan 1-forms
\p{cforms}, it is easy to find that the covariant constraints
which express all spinor covariant
derivatives of bosonic superfields in terms of the Goldstone fermions
amount to setting equal to zero
the spinor projections of these 1-forms
(these conditions are a particular case of the inverse Higgs effect \cite{InvH}).
Thus, in the case at hand we impose the following constraints:
\be\label{hyper1}
\left. \omega_D=\omega_V\left|=\bar\omega_V\right|=\omega_{V_3}\right|=0 \;,
\ee
where $|$ means restriction to spinor projections.
These constraints are manifestly covariant
under the whole supergroup $D(2,1;\alpha)$. They allow
one to express the Goldstone spinor superfields
as the spinor derivatives of the residual bosonic Goldstone superfields
$u, \Lambda, \bar\Lambda, \phi$ and  imply
some irreducibility constraints for the latter
\bea\label{hyper2}
&& D^1 \Lambda = -2i\alpha \Lambda \left( \bpsi^1+\Lambda\bpsi^2\right),\;
D^1\bLam=-2i\alpha\left(\bpsi^2-\bLam\bpsi^1\right),\;
D^1 \phi =-2\alpha\left( \bpsi^1+\Lambda\bpsi^2\right),\nn
&& D^2\Lambda =2i\alpha\left(\bpsi^1+\Lambda\bpsi^2\right), \;
D^2\bLam=-2i\alpha\bLam\left( \bpsi^2-\bLam\bpsi^1\right),\;
D^2 \phi = 2\alpha \left(\bpsi^2-\bLam \bpsi^1\right), \nn
&& D^1 u =2i \bpsi^1 ,\;  D^2 u = 2i\bpsi^2, \quad \dot{u}=2z
\eea
(and c.c.). The irreducibility conditions, in this and other cases which
we shall consider further, arise due
to the property that the Goldstone fermionic superfields are simultaneously expressed
by \p{hyper2} in terms of spinor derivatives of different bosonic superfields. Then,
eliminating these spinor superfields, we end up with the relations between the spinor
derivatives of bosonic Goldstone superfields. In order to make the use of these constraints the
most feasible, it is advantageous to pass to the new variables
\be\label{hyper3}
q^1=\frac{e^{\frac{1}{2}(\alpha u -i\phi)}}{\sqrt{1+\Lambda\bLam}}\Lambda ,\quad
q^2=-\frac{e^{\frac{1}{2}(\alpha u -i\phi)}}{\sqrt{1+\Lambda\bLam}},\quad
{\bar q}_1=\frac{e^{\frac{1}{2}(\alpha u +i\phi)}}{\sqrt{1+\Lambda\bLam}}\bLam ,\;
{\bar q}_2=-\frac{e^{\frac{1}{2}(\alpha u +i\phi)}}{\sqrt{1+\Lambda\bLam}}\,.
\ee
In terms of these variables the irreducibility constraints acquire the manifestly
$SU(2)$ covariant form
\be\label{hyper4}
D^{(i}q^{j)}=0,\quad \bD{}^{(i}q^{j)}=0\,.
\ee
This is just the $N=4,d=1$ hypermultiplet multiplet we considered in the previous section.

The rest of the supermultiplets from Table 1 may be obtained similarly. For example, the
tensor $N=4,d=1$ supermultiplet corresponds to the coset with the $V_3$ generator in the
stability subgroup and so on. The detailed discussion of all cases may be found in \cite{ikl1}.

\section{N=8 Supersymmetry}\label{sec:3}

Most of the models explored in the previous sections possess
$N{\leq}4, d{=}1$ supersymmetries. Much less is known
about higher-$N$ systems. Some of them were addressed many years
ago in the seminal paper \cite{CR}, within an on-shell Hamiltonian approach.
Some others (with $N{=}8$) received attention lately \cite{DE, BMZ}.

As we stressed many times, the natural formalism for dealing with supersymmetric models is the
off-shell superfield approach. Thus, for the construction of new SQM models
with extended $d{=}1$ supersymmetry, one needs, first of all, the
complete list of the corresponding off-shell $d{=}1$ supermultiplets
and the superfields which encompass these multiplets. One of
the peculiarities of $d{=}1$ supersymmetry is that some
of its off-shell multiplets cannot be obtained via a direct
dimensional reduction from the multiplets of higher-$d$ supersymmetries
with the same number of spinorial charges. Another peculiarity is that
some on-shell multiplets
of the latter have {\it off-shell\/} $d{=}1$ counterparts.

In the previous section we considered nonlinear realizations of the
finite-dimensional $N=4$ superconformal group in $d{=}1$. We showed that
the irreducible superfields representing one or another off-shell $N=4, d{=}1$
supermultiplet come out as Goldstone superfields parameterizing one
or another coset manifold of the superconformal group.
The superfield irreducibility constraints naturally emerge as a part
of manifestly covariant inverse Higgs \cite{InvH} conditions on the relevant
Cartan superforms.

This method is advantageous in that it automatically specifies the
superconformal properties of the involved supermultiplets,  which
are of importance. The application of the nonlinear realization approach
to the  case of $N{=}8, d{=}1$ supersymmetry was
initiated in \cite{bikl2}. There, nonlinear realizations of
the $N{=}8, d{=}1$ superconformal group $OSp(4^\star\vert 4)$ in its two
different cosets were considered, and it was shown that two interesting $N{=}8, d{=}1$
multiplets, with off-shell field contents ({\bf 3, 8, 5})
and ({\bf 5, 8, 3}),
naturally come out as the corresponding Goldstone multiplets. These supermultiplets admit a few
inequivalent splittings into pairs of irreducible off-shell $N{=}4, d{=}1$
multiplets, such that different $N{=}4$ superconformal subgroups of
$OSp(4^\star\vert 4)$, viz. $SU(1,1\vert 2)$ and $OSp(4^\star\vert 2)$,
are manifest for different splittings. Respectively, the
off-shell component action of the given $N{=}8$ multiplet in general
admits several different representations in terms of $N{=}4, d{=}1$ superfields.

Now we are going to present a superfield description
of all other linear off-shell $N{=}8, d{=}1$ supermultiplets with ${\bf 8}$
fermions, in both $N{=}8$ and $N{=}4$ superspaces.

Towards deriving an exhaustive list of off-shell $N{=}8$ supermultiplets and
the relevant constrained $N{=}8, d{=}1$ superfields, we could proceed in
the same way as in the case of $N{=}4$ supermultiplets, i.e. by considering nonlinear realizations of all known
$N{=}8$ superconformal groups in their various cosets. However,
this task is more complicated, as compared to the $N{=}4$ case,
in view of the existence of many inequivalent
$N{=}8$ superconformal groups ($OSp(4^\star\vert 4)$, $OSp(8\vert 2)$, $F(4)$
and $SU(1, 1\vert 4)$, see e.g. \cite{VP1}), with numerous coset manifolds.

In order to avoid these complications, we take advantage of two fortunate
circumstances.  Firstly, as we already know,
the field contents of {\it linear\/} off-shell
multiplets of $N{=}8, d{=}1$ supersymmetry with {\bf 8} physical fermions
range from  ({\bf 8, 8, 0}) to ({\bf 0, 8, 8}), with the intermediate
multiplets corresponding to all possible splittings of ${\bf 8}$
bosonic fields into physical and auxiliary ones. Thus we are aware of
the full list of such multiplets, independently of the issue
of their interpretation as the Goldstone ones parameterizing the
proper superconformal cosets.

The second circumstance allowing us to advance without resorting to
the nonlinear realizations techniques is the aforesaid
existence of various splittings of $N{=}8$ multiplets into pairs of irreducible
$N{=}4$ supermultiplets. We know how to represent the latter in terms
of constrained $N{=}4$ superfields, so it proves to be a matter of simple
algebra to guess the form of the four extra supersymmetries mixing
the $N{=}4$ superfields inside each pair and extending the manifest
$N{=}4$ supersymmetry to $N{=}8$. After fixing such pairs, it is again
rather easy to embed them into appropriately constrained $N{=}8, d{=}1$
superfields.

\subsection{N=8, d=1 superspace}
The maximal automorphism group of $N{=}8, d{=}1$ super Poincar\'e algebra
(without central charges) is $SO(8)$ and so eight real Grassmann coordinates
of $N{=}8, d{=}1$ superspace $\mathbb{R}^{(1|8)}$ can be arranged into one of three 8-dimensional
real irreps of $SO(8)$. The constraints defining the irreducible
$N{=}8$ supermultiplets in general break this $SO(8)$ symmetry. So,
it is preferable to split the
8 coordinates into two real quartets
\be\label{def1}
\mathbb{R}^{(1|8)} = (t, \,\theta_{ia},\, \vt_{\alpha A})\,,\qquad\quad
\overline{ \left( \theta_{ia}\right)}=\theta^{ia},\;
\overline{ \left( \vt_{\alpha A}\right)}=\vt^{\alpha A},
\quad i,a,\alpha, A= 1,2\,,
\ee
in terms of which only four commuting automorphism $SU(2)$ groups will be explicit.
The further symmetry breaking can be understood as the identification of some of these $SU(2)$, while
extra symmetries, if they exist, mix different $SU(2)$ indices.
The corresponding covariant derivatives are defined by
\be\label{def2}
D^{ia}=\frac{\partial}{\partial\theta_{ia}}+i\theta^{ia} \partial_t\; , \;
\nabla^{\alpha A}=\frac{\partial}{\partial\vt_{\alpha A}}
+i\vt^{\alpha A} \partial_t\;.
\ee
By construction, they obey the algebra
\be\label{def3}
\left\{ D^{ia}, D^{jb}\right\} =2i \epsilon^{ij}\epsilon^{ab}\partial_t\,,\quad
\left\{ \nabla^{\alpha A}, \nabla^{\beta B}\right\}
=2i \epsilon^{\alpha\beta}\epsilon^{AB}\partial_t \; .
\ee

\subsection{N=8, d=1 supermultiplets}
As we already mentioned, our real strategy of deducing a superfield
description of the $N{=}8, d{=}1$ supermultiplets
consisted in selecting an appropriate pair of constrained $N{=}4, d{=}1$
superfields  and then guessing the constrained
$N{=}8$ superfield.  Now, just to make the presentation more
coherent, we turn the argument around and start with postulating the $N{=}8, d{=}1$
constraints. The $N{=}4$ superfield formulations will be deduced from the $N{=}8$ ones.

\subsubsection{Supermultiplet ({\bf 0, 8, 8})}

The off-shell $N{=}8, d{=}1$ supermultiplet ({\bf 0, 8, 8}) is carried out by two real
fermionic $N{=}8$ superfields $\Psi^{aA}, \Xi^{i\alpha}$ subjected to the following
constraints:
\bea
&& D^{(i a}\Xi^{j)}_{\alpha}=0,\; D^{i(a}\Psi^{b)}_{A}=0,\quad
\nabla^{(\alpha A}\Xi^{\beta)}_{i}=0,\;
\nabla^{\alpha( A}\Psi^{B)}_{a}=0, \label{1con1} \\
&& \nabla^{\alpha A}\Psi_A^a=D^{ia}\Xi_i^\alpha,\quad
\nabla^{\alpha A}\Xi_\alpha^i=-D^{ia}\Psi_a^A\;. \label{1con2}
\eea

In order to understand the structure of this supermultiplet in terms of $N{=}4$
superfields
we proceed as follows. As a first step, let us single out
the $N{=}4$ subspace in the $N{=}8$ superspace $\mathbb{R}^{(1|8)}$ as the set of coordinates
\be
\mathbb{R}^{(1|4)} = \left( t, \theta_{ia} \right) \subset \mathbb{R}^{(1|8)}, \label{ss1}
\ee
and expand the $N{=}8$ superfields over the
extra Grassmann coordinate $\vartheta_{\alpha A}$. Then we observe
that the constraints \p{1con2} imply that the spinor derivatives
of all involved superfields with respect to $\vartheta_{\alpha A}$
can be expressed in terms of
spinor derivatives with respect to $\theta_{i a}\,$. This means that
the only essential $N{=}4$ superfield components of $\Psi^{aA}$ and
$\Xi^{i\alpha}$ in their $\vartheta$-expansion are the first ones
\be\label{1comp1}
\psi^{aA} \equiv \Psi^{aA}|_{\vt=0}\;,\quad \xi^{i\alpha} \equiv
\Xi^{i\alpha}|_{\vt=0}\,.
\ee

These fermionic $N{=}4$ superfields are subjected, in virtue of
eqs. \p{1con1} and \p{1con2}, to the irreducibility constraints in $N{=}4$
superspace
\be\label{1con3}
D^{a(i} \xi^{j)\alpha}=0,\;
D^{i(a} \psi^{b)A}=0\,.
\ee
As it follows from Section~2, these superfields are just
two fermionic $N{=}4$ hypermultiplets, each carrying ({\bf 0, 4, 4})
independent component fields. So, being combined together, they accommodate
the whole off-shell component content of the  $N{=}8$
multiplet ({\bf 0, 8, 8}), which proves that the $N{=}8$ constraints
\p{1con1}, \p{1con2} are the true choice.

Thus, from the $N{=}4$ superspace perspective, the
$N{=}8$ supermultiplet ({\bf 0, 8, 8}) amounts to the sum of two $N{=}4, d{=}1$
fermionic hypermultiplets  with the
off-shell component content ({\bf 0, 4, 4}) $\oplus$ ({\bf 0, 4, 4}).

The transformations of the implicit $N{=}4$ Poincar\'e supersymmetry,
completing the manifest one to the full $N{=}8$ supersymmetry, have the
following form in terms of the $N{=}4$ superfields defined above:
\be\label{1tr1}
\delta\psi^{aA}=\frac{1}{2}\eta^{A\alpha} D^{ia}\xi_{i\alpha},\quad
\delta \xi^{i\alpha}=-\frac{1}{2}\eta^\alpha_A D^i_a \psi^{aA}.
\ee

The invariant free action can be written as \be\label{1action}
S=\int dt d^4\theta \left[ \theta^{ia}\theta_i^b \psi_a^A
\psi_{bA}+\theta^{ia}\theta_a^j \xi^\alpha_i \xi_{j\alpha}\right].
\ee

Because of the presence of explicit theta's in the action
\p{1action}, the latter
is not manifestly invariant even with respect to the manifest $N{=}4$
supersymmetry. Nevertheless, one can check that \p{1action}
is invariant under this supersymmetry, which is realized on the
superfields as
\be\label{1transf3}
\delta^* \psi^{aA}=-\varepsilon_{jb}Q^{jb} \psi^{aA},\;
\delta^* \xi^{i\alpha}=-\varepsilon_{jb}Q^{jb} \xi^{i\alpha},
\ee
where
\be
Q^{ia}=\frac{\partial}{\partial\theta_{ia}}-i\theta^{ia} \partial_t\;,
\ee
$\varepsilon_{ia}$ is the supertranslation parameter and
${}^*$ denotes the `active' variation (taken at a fixed point of
the $N{=}4$ superspace).

\subsubsection{Supermultiplet ({\bf 1, 8, 7})}
This supermultiplet can be described by a single scalar $N{=}8$ superfield
$\cal U$ which obeys the following irreducibility conditions:
\bea
&& D^{ia}D_a^j{\cal U}=-\nabla^{\alpha j}\nabla_{\alpha}^i{\cal U},
\label{2con1a} \\
&& \nabla^{(\alpha i}\nabla^{\beta)j}{\cal U}=0,\quad
D^{i(a}D^{j b)}{\cal U} =0 \;. \label{2con1b}
\eea
Let us note that the constraints \p{2con1a} reduce the manifest
R-symmetry to $[SU(2)]^3$, due to the identification
of the indices $i$ and $A$ of the covariant derivatives
$D^{ia}$ and $\nabla^{\alpha A}$.

This supermultiplet possesses a unique decomposition
into the pair of $N{=}4$ supermultiplets as
({\bf 1, 8, 7}) = ({\bf 1, 4, 3}) $\oplus$ ({\bf 0, 4, 4}). The corresponding
$N{=}4$ superfield projections  can be defined as
\be\label{2defcomp}
u={\cal U}|_{\vt=0} ,\quad \psi^{i\alpha}=\nabla^{\alpha i}{\cal U}|_{\vt=0}\;,
\ee
and they obey the standard constraints
\be\label{2con2}
D^{(i a}\psi^{j)\alpha}=0\;, \quad D^{i(a}D^{j b)} u=  0\;.
\ee
The second constraint directly follows from \p{2con1b}, while the first one is implied
by the relation
\be\label{2con1c}
\frac{\partial}{\partial t} D^{(i}_a\nabla^{j)}_{\alpha}{\cal U} =0\,,
\ee
which can be proven by applying the differential operator $D^{kb}\nabla^{\beta l}$
to the $N{=}8$ superfield constraint \p{2con1a} and making use of the algebra of covariant derivatives.

The  additional implicit $N{=}4$ supersymmetry is realized on these
$N{=}4$ superfields as follows:
\be\label{2tr}
\delta u=-\eta_{i\alpha}\psi^{i\alpha},\quad \delta \psi^{i\alpha}
=-\frac{1}{2}\eta^\alpha_j D^{ia} D^j_a u \;.
\ee
The simplest way to deal with the action for this supermultiplet
is to use harmonic superspace \cite{{harm},{book},{IL}}, but this approach is
out of the scope of the present Lectures.

\subsubsection{Supermultiplet ({\bf 2, 8, 6})}
The $N{=}8$ superfield formulation of  this supermultiplet involves
two scalar bosonic superfields ${\cal U}, \Phi$ obeying
the constraints
\bea
&& \nabla^{(a i}\nabla^{b)j}{\cal U}=0, \quad
\nabla^{a(i}\nabla^{b j)}\Phi=0 , \label{31} \\
&& \nabla^{ai}{\cal U}=D^{ia}\Phi, \quad \nabla^{ai}\Phi=-D^{ia}{\cal U}
\label{32}
\eea
where we have identified the indices $i$ and $A$, $a$ and $\alpha$ of the
covariant derivatives, thus retaining only two manifest SU(2) automorphism
groups.  From \p{31}, \p{32} some useful corollaries follow:
\bea
&& D^{ia}D^j_a{\cal U}+\nabla^{aj}\nabla^i_a{\cal U}=0,
\quad D^{i(a}D^{j b)}{\cal U}=0, \label{33} \\
&& D^{ia}D_i^b \Phi+\nabla^{b i}\nabla_i^a \Phi=0, \quad
D^{(i a}D^{j)b}\Phi=0.\label{34}
\eea
Comparing \p{33}, \p{34} and \p{31} with \p{2con1a}, \p{2con1b},
we observe that the $N{=}8$ supermultiplet with the field content
({\bf 2, 8, 6}) can be obtained by combining  two ({\bf 1, 8, 7})
supermultiplets and imposing the additional
relations \p{32} on the corresponding $N{=}8$ superfields.

In order to construct the invariant actions
and prove that the above $N{=}8$ constraints indeed yield
the multiplet ({\bf 2, 8, 6}), we should reveal the structure
of this supermultiplet in terms of $N{=}4$ superfields,
as we did in the previous cases. However, in the case at hand,
we have two different choices for splitting the ({\bf 2, 8, 6})
supermultiplet
\begin{itemize}
\item {\bf 1. } ({\bf 2, 8, 6}) = ({\bf 1, 4, 3}) $\oplus$ ({\bf 1, 4, 3})
\item {\bf 2. } ({\bf 2, 8, 6}) = ({\bf 2, 4, 2}) $\oplus$ ({\bf 0, 4, 4})
\end{itemize}
As already mentioned, the possibility to have a few  different
off-shell $N{=}4$ decompositions of the same $N{=}8$ multiplet
 is related to different choices of the manifest $N{=}4$
supersymmetries, as subgroups of the $N{=}8$ super Poincar\'e group.
We shall treat both options.

\noindent{\bf 1. } ({\bf 2, 8, 6}) = ({\bf 1, 4, 3}) $\oplus$ ({\bf 1, 4, 3})\\

In order to describe the $N{=}8$ ({\bf 2, 8, 6}) multiplet in terms of
$N{=}4$ superfields, we should choose the appropriate $N{=}4$ superspace.
The first (evident) possibility is to choose the $N{=}4$ superspace
with coordinates $( t, \theta_{ia} )$. In this superspace
one $N{=}4$ Poincar\'e supergroup is naturally realized, while the second
one mixes two irreducible $N{=}4$
superfields which comprise the  $N{=}8$ ({\bf 2, 8, 6}) supermultiplet
in question. Expanding the $N{=}8$ superfields ${\cal U},\Phi$
in $\vt^{ia}$,  one finds that the constraints \p{31}, \p{32} leave
in ${\cal U}$ and $\Phi$ as independent $N{=}4$ projections
only those of zeroth order in $\vt^{ia}$
\be\label{4def1}
\left. u={\cal U}\right|_{\vt_{i\alpha}=0}\,,\quad \left.
\phi=\Phi\right|_{\vt_{i\alpha}=0}\,.
\ee
Each $N{=}4$ superfield proves to be subjected, in virtue
of \p{31}, \p{32}, to the additional constraint
\be\label{4aconn4}
D^{i(a}D^{j b)}u=0\,, \quad D^{(i a}D^{j)b}\phi=0\,.
\ee
Thus we conclude that our $N{=}8$  multiplet ${\cal U}, \Phi$, when
rewritten in terms of $N{=}4$ superfields, amounts to a direct sum
of two  $N{=}4$ multiplets $u$ and $\phi$, both having the same
off-shell field contents ({\bf 1, 4, 3}).

The transformations of the implicit $N{=}4$ Poincar\'e supersymmetry,
completing the manifest
one to the full $N{=}8$ Poincar\'e supersymmetry, have the following form
in terms of these $N{=}4$ superfields:
\be\label{4transf1}
\delta^* u= - \eta_{ia}D^{ia}\phi, \quad \delta^* \phi= \eta_{ia}D^{ia}u \;.
\ee

It is rather easy to construct the action in terms of $N{=}4$ superfields
$u$ and $\phi$, such that it is  invariant with respect
to the implicit $N{=}4$ supersymmetry \p{4transf1}. The generic action has the form
\be\label{4aaction1}
S= \int dt d^4\theta {\cal F}(u,\phi)\;,
\ee
where the function ${\cal F}$ obeys the Laplace equation
\be\label{4alaplace}
{\cal F}_{uu}+{\cal F}_{\phi\phi}=0 \;.
\ee

\noindent{\bf 2. } ({\bf 2, 8, 6}) = ({\bf 2, 4, 2}) $\oplus$ ({\bf 0, 4, 4})\\

There is a more sophisticated choice of a $N{=}4$ subspace in the
$N{=}8, d{=}1$ superspace, which gives rise to the second possible $N{=}4$
superfield splitting of the considered $N{=}8$ supermultiplet, that is
into the multiplets ({\bf 2, 4, 2}) and ({\bf 0, 4, 4}).

First of all, let us define a new set of covariant derivatives
\be\label{4bder}
\cD^{ia}=\frac{1}{\sqrt{2}}\left( D^{ia}-i\nabla^{ai}\right),\quad
\cbD^{ia}=\frac{1}{\sqrt{2}}\left( D^{ia}+i\nabla^{ai}\right),\quad
\left\{ \cD^{ia},\cbD^{jb}\right\}=2i\epsilon^{ij}\epsilon^{ab}\partial_t\,,
\ee
and new $N{=}8$ superfields $\cV, \cbV$ related to the original ones as
\be\label{4dsf}
\cV={\cal U}+i\Phi, \quad \cbV={\cal U}-i\Phi\;.
\ee
In this basis the constraints \p{31}, \p{32} read
\bea
&& \cD^{ia}\cV=0 ,\quad \cbD^{ia}\cbV=0, \nn
&& \cD^{i(a}\cD^{jb)}\cbV+\cbD^{i(a}\cbD^{j b)}\cV=0,\quad
\cD^{(i a}\cD^{j)b}\cbV-\cbD^{(i a}\cbD^{j)b}\cV=0. \label{4bcon1}
\eea
Now we split the complex quartet covariant derivatives \p{4bder}
into two sets of the doublet $N{=}4$ ones as
\be\label{4bder1}
D^i=\cD^{i1},\; \bD^{i}=\cbD^{i2},\quad \nabla^i=\cD^{i2},\;
\bar\nabla{}^i=-\cbD^{i1}
\ee
and cast the constraints \p{4bcon1} in the form
\bea
&& D^i\cV=0,\; \nabla^i\cV=0,\quad \bD_i\cbV=0,\; \overline\nabla_i\cbV=0 ,\nn
&& D^iD_i \cbV-{\overline\nabla}_i\overline\nabla{}^i\cV=0,
\quad D^i\nabla^j\cbV-\bD{}^i\overline\nabla{}^j\cV=0 \;. \label{4bcon2}
\eea
Next, as an alternative $N{=}4$ superspace, we choose the set of coordinates
closed under the action of $D^i, \bD{}^i$, i.e.
\be
\left( t\,,\; \theta_{i1} + i\vartheta_{i1}\,,\; \theta_{i2}
- i \vartheta_{i2} \right), \label{4bss}
\ee
while the $N{=}8$ superfields are expanded with respect to the
orthogonal combinations $\theta_{i1} - i \vartheta_{i1}$,
$\theta_{i2} + i\vartheta_{i2}$ which are annihilated by $D^i, \bD{}^i$.

As a consequence of the constraints \p{4bcon2}, the quadratic action
of the derivatives $\nabla^i$ and $\overline{\nabla}{}^i$ on every $N{=}8$
superfield $\cV, \cbV$  can be expressed as $D^i, \bD{}^i$
of some other superfield. Therefore, only
the zeroth and first order  components
of each $N{=}8$ superfield are independent
$N{=}4$ superfield projections. Thus, we are left with the following set
of $N{=}4$ superfields:
\be\label{4bn4sf}
\left. v=\cV\right|,\; \left. {\bar v}=\cbV\right|,\quad
\left. \psi^i=\overline\nabla{}^i\cV\right| ,\; \left.
\bpsi^i=-\nabla^i\cbV\right| \;.
\ee
These $N{=}4$ superfields prove to be subjected to the
additional constraints which also follow from \p{4bcon2}
\be\label{4bcon3}
D^i v=0\,,\;\; \bD{}^i {\bar v}=0\,,\quad D^i\psi^j=0\,,\;\;
\bD{}^i\bpsi{}^j=0\,,\;\; D^i\bpsi{}^j=-\bD{}^i\psi^j \;.
\ee
The $N{=}4$ superfields $v, {\bar v}$ comprise the standard
$N{=}4, d{=}1$ chiral multiplet ({\bf 2, 4, 2}),
while the $N{=}4$ superfields $\psi^i, \bpsi^j$ subjected to
\p{4bcon3} and both having the off-shell contents ({\bf 0, 4, 4})
are recognized as the fermionic version of the $N{=}4, d{=}1$
hypermultiplet.

The implicit $N{=}4$ supersymmetry is realized by the transformations
\bea\label{4btr}
&& \delta v=-\bar\eta{}^i \psi_i,\quad \delta\psi^i
=-\frac{1}{2}\bar\eta{}^i D^2 {\bar v}-2i\eta^i{\dot v}, \nn
&& \delta {\bar v}= \eta_i \bpsi^i, \quad \delta\bpsi_i
=\frac{1}{2}\eta_i \bD^2 v +2i\bar\eta_i{\dot {\bar v}}\:.
\eea

The invariant free action has the following form:
\be\label{4baction}
S_f=\int dt d^4\theta v{\bar v} -\frac{1}{2}\int dt d^2{\bar\theta}
\psi^i\psi_i-\frac{1}{2}\int dt d^2\theta \bpsi_i\bpsi{}^i\;.
\ee
Let us note that this very simple form of the action for the $N{=}4$ ({\bf 0, 4, 4})
supermultiplet $\psi_i,\bpsi{}^j$ is related to our choice of
the $N{=}4$ superspace.
It is worthwhile to emphasize that all differently looking superspace off-shell actions of the
multiplet ({\bf 0, 4, 4}) yield the same component action for
this multiplet.

\subsubsection{Supermultiplet ({\bf 3, 8, 5})}

In the $N{=}8$ superspace this supermultiplet is described
by the triplet of bosonic superfields
$\cV^{ij}$ obeying the irreducibility constraints
\be\label{5con}
D_a^{(i}\cV^{jk)} =0 \; , \quad \nabla_\alpha{}^{(i}\cV^{jk)} =0 \; .
\ee
So, three out of four original automorphism $SU(2)$ symmetries
remain manifest in this description.

The $N{=}8$ supermultiplet ({\bf 3, 8, 5}) can be decomposed into
$N{=}4$ supermultiplets in two ways
\begin{itemize}
\item {\bf 1. } ({\bf 3, 8, 5}) = ({\bf 3, 4, 1}) $\oplus$ ({\bf 0, 4, 4})
\item {\bf 2. } ({\bf 3, 8, 5}) = ({\bf 1, 4, 3}) $\oplus$ ({\bf 2, 4, 2})
\end{itemize}

As in the previous case, we discuss both options.\vspace{0.5cm}

\noindent{\bf 1. } ({\bf 3, 8, 5}) = ({\bf 3, 4, 1}) $\oplus$ ({\bf 0, 4, 4})\\

This  splitting requires choosing the coordinate set \p{ss1}
as the relevant $N{=}4$ superspace. Expanding the $N{=}8$ superfields $\cV^{ij}$
in $\vt_{i\alpha}$,  one finds that the constraints \p{5con} leave
in $\cV^{ij}$ the following four bosonic and four fermionic $N{=}4$ projections:
\be\label{5acomp}
\left. v^{ij}=\cV^{ij}\right| ,\quad \left. \xi^i_\alpha \equiv
\nabla_{j\alpha}\cV^{ij}\right|,\quad
\left. A\equiv \nabla^\alpha_i \nabla_{j\alpha}\cV^{ij}\right|
\ee
where $|$ means the restriction to $\vt_{i\alpha}=0$.
As a consequence of \p{5con}, these $N{=}4$ superfields
obey the constraints
\bea
&& D_a^{(i}v^{jk)}=0\,,\quad  D_a^{(i}\xi^{j)}_\alpha =0\,,\nn
&& A= 6m -D^a_i D_{aj} v^{ij}\,, \;m = \mbox{const}\, . \label{5acon1}
\eea
Thus, for the considered splitting, the $N{=}8$ tensor multiplet
superfield  $\cV^{ij}$ amounts to a direct sum of the $N{=}4$ `tensor' multiplet
superfield $v^{ij}$ with the  off-shell content $({\bf 3,\, 4,\, 1})$
and a fermionic $N{=}4$ hypermultiplet $\xi^i_\alpha$
with the off-shell content ({\bf 0, 4, 4}), plus a constant $m$
of the mass dimension.

\noindent{\bf 2. } ({\bf 3, 8, 5}) = ({\bf 1, 4, 3}) $\oplus$ ({\bf 2, 4, 2})\\

This option corresponds to another choice of $N{=}4$ superspace,
which amounts to dividing
the $N{=}8, d{=}1$ Grassmann coordinates into doublets, with respect to some other
$SU(2)$ indices.
The relevant splitting of $N{=}8$ superspace into the $N{=}4$ subspace
and the complement of the latter can be performed  as follows. Firstly,
we define the new covariant derivatives  as
\bea\label{5bder}
&&  D^a\equiv \frac{1}{\sqrt{2}}\left( D^{1a}+i\nabla^{a1}\right),\;
\bD_a\equiv \frac{1}{\sqrt{2}}\left( D_a^{2}-i\nabla_a^{2}\right), \nn
&& \nabla^a\equiv \frac{i}{\sqrt{2}}\left( D^{2a}+i\nabla^{a2}\right),\;
\bar\nabla_a\equiv \frac{i}{\sqrt{2}}\left( D_a^{1}-i\nabla_a^{1}\right).
\eea
Then we choose the set of coordinates
closed under the action of $D^a, \bar D_a$, i.e.
\be
\left( t\,,\; \theta_{1a} - i\vartheta_{a1}\,,\;
\theta^{1a} + i \vartheta^{a1} \right), \label{ss2}
\ee
while the $N{=}8$ superfields are expanded with respect to
the orthogonal combinations $\theta^a_2 - i \vartheta^a_2\,$,
$\theta^a_1 + i\vartheta^a_1$ annihilated by $D^a, \bar D_a$.

The basic constraints \p{5con}, being rewritten in the basis \p{5bder},
take the form
\bea\label{5bcon1}
&& D^a\varphi=0\,, \quad D^a v -\nabla^a \varphi=0\,,
\quad \nabla^a v + D^a{\bar\varphi}=0\,,\quad \nabla^a \bar\varphi =0\,, \nn
&& {\overline{\nabla}}_a \varphi=0,\quad
\overline{\nabla}_a v+\bD_a \varphi=0\,,\quad
\bD_a v - \overline{\nabla}_a \bar\varphi =0\,,
\quad \bD_a\bar\varphi=0
\eea
where
\be\label{5bsf}
v\equiv -2i \cV^{12}\,,\quad \varphi \equiv \cV^{11}\,,\quad
\bar\varphi\equiv \cV^{22}\,.
\ee
Due to the constraints \p{5bcon1}, the derivatives $\nabla^a$
and $\overline{\nabla}_a$ of every $N{=}8$
superfield in the triplet $\left(\cV^{12}, \cV^{11}, \cV^{22} \right)$ can be
expressed as $D^a, \bD_a$ of some other superfield. Therefore, only
the zeroth order (i.e. taken at $\theta^a_2-i\vt_2^a=\theta^a_1+i\vt^a_1=0$)
components of each $N{=}8$ superfield are independent
$N{=}4$ superfield projections. These $N{=}4$ superfields are subjected
to the additional constraints which also follow from \p{5bcon1}
\be\label{5bcon2}
D^aD_a v=\bD_a \bD^a v=0\,,\quad D^a\varphi=0\,,\; \bD_a \bar\varphi=0\,.
\ee
The $N{=}4$ superfields $\varphi,\bar\varphi$ comprise the standard
$N{=}4, d{=}1$ chiral multiplet ({\bf 2, 4, 2}),
while the $N{=}4$ superfield $v$ subjected to \p{5bcon2} has the
needed off-shell content ({\bf 1, 4, 3}).

The implicit $N{=}4$ supersymmetry acts on the $N{=}4$ superfields
$v,\varphi,\bar\varphi$ as follows:
\be\label{5btransf3}
\delta^* v=\eta_a D^a \bar\varphi
+\bar\eta{}^a\bD_a \varphi\,,\quad \delta^*\varphi=-\eta_a D^a v\,,\quad
\delta^*\bar\varphi =-\bar\eta{}^a\bD_a v\,.
\ee

Invariant $N{=}4$ superfield actions for both decompositions of the
$N{=}8$ multiplet ({\bf 3, 8, 5}) were presented in Ref. \cite{bikl2}.

\subsection{Supermultiplet ({\bf 4, 8, 4})}

This supermultiplet can be described by a quartet of $N{=}8$ superfields
${\cal Q}^{a\alpha}$ obeying the constraints
\be\label{6con}
D^{(a}_i{\cal Q}^{b)\alpha}=0,\quad \nabla^{(\alpha}_i{\cal Q}^{\beta)}_a=0 \;.
\ee
Let us note that the constraints \p{6con} are manifestly
covariant with respect to three $SU(2)$ subgroups realized
on the indices $i,a$ and $\alpha$.

{}From \p{6con} some important relations follow:
\be\label{6rel}
D^{ia}D^{jb}{\cal Q}^{c\alpha}
=2i\epsilon^{ij}\epsilon^{cb}{\dot{\cal Q}}{}^{a\alpha},\quad
\nabla^{i\alpha}\nabla^{j\beta}{\cal Q}^{a\gamma}
=2i\epsilon^{ij}\epsilon^{\gamma\beta}{\dot{\cal Q}}{}^{a\alpha}\;.
\ee
Using them, it is possible to show that the superfields
${\cal Q}^{a\alpha}$ contain the following independent components:
\be\label{6comp1}
\left. {\cal Q}^{a\alpha}\right|,\quad \left. D^i_a {\cal Q}^{a\alpha}\right|,
\quad \left.
\nabla^i_\alpha {\cal Q}^{a\alpha}\right|,\quad
\left. D^i_a \nabla^j_\alpha {\cal Q}^{a\alpha}\right|,
\ee
where $|$ means now restriction to $\theta_{ia}=\vt_{i\alpha}=0$.
This directly proves that we deal with the
irreducible ({\bf 4, 8, 4}) supermultiplet.

There are three different possibilities to split this $N{=}8$ multiplet
into the $N{=}4$ ones

\begin{itemize}
\item {\bf 1. } ({\bf 4, 8, 4}) = ({\bf 4, 4, 0}) $\oplus$ ({\bf 0, 4, 4})
\item {\bf 2. } ({\bf 4, 8, 4}) = ({\bf 3, 4, 1}) $\oplus$ ({\bf 1, 4, 3})
\item {\bf 3. } ({\bf 4, 8, 4}) = ({\bf 2, 4, 2}) $\oplus$ ({\bf 2, 4, 2})
\end{itemize}

Once again, we shall consider all three cases separately.\vspace{0.5cm}

\noindent{\bf 1. } ({\bf 4, 8, 4}) = ({\bf 4, 4, 0}) $\oplus$ ({\bf 0, 4, 4})\\

This case implies the choice of the $N{=}4$ superspace \p{ss1}.
Expanding the $N{=}8$ superfields ${\cal Q}^{a\alpha}$
in $\vt_{i\alpha}$,  one may easily see that the constraints \p{6con}
leave in ${\cal Q}^{a\alpha}$ the following four bosonic and four
fermionic $N{=}4$ superfield projections:
\be\label{6acomp}
\left. q^{a\alpha}={\cal Q}^{a\alpha}\right| ,\quad
\left. \psi^{ia} \equiv \nabla^i_\alpha {\cal Q}^{a\alpha}\right|\,.
\ee
Each $N{=}4$ superfield is subjected, in virtue of \p{6con}, to an additional constraint
\be
 D^{i(a} q^{b)\alpha}=0\,,\quad  D^{i(a}\psi^{b)i} =0 . \label{6acon1}
\ee
Consulting Section~2, we come to the conclusion that these are just
the hypermultiplet $q^{i\alpha}$ with the off-shell field content ({\bf 4, 4, 0})
and a fermionic analog of the $N{=}4$ hypermultiplet $\psi^{ia}$ with the
field content ({\bf 0, 4, 4}).

The transformations of the implicit $N{=}4$ Poincar\'e supersymmetry
have the following form in terms of these $N{=}4$ superfields:
\be\label{6atransf1}
\delta^* q^{a\alpha}=\frac{1}{2}\eta^{i\alpha}\psi^a_i\,,\quad
\delta^* \psi^{ia}=-2i\eta^{i\alpha}{\dot q}{}^{a}_a\;.
\ee

\noindent{\bf 2. } ({\bf 4, 8, 4}) = ({\bf 3, 4, 1}) $\oplus$ ({\bf 1, 4, 3})\\

In order to describe this $N{=}4$ superfield realization of the
$N{=}8$ supermultiplet $({\bf 4,\, 8,\, 4})$, we introduce the $N{=}8$
superfields ${\cal V}{}^{ab}, {\cal V}$ as
\be\label{6csf}
{\cal Q}{}^{a\alpha} \equiv \delta^\alpha_b{\cal V}{}^{a b}
-\epsilon^{a\alpha}{\cal V}\,, \quad {\cal V}{}^{a b} = {\cal V}{}^{b a}\,,
\ee
and use the covariant derivatives \p{4bder} to rewrite
the basic constraints \p{6con} as
\bea
&& \cD^{i(a}{\cal V}^{bc)}=0,
\quad \cbD{}^{i(a}{\cal V}^{bc)}=0, \label{6ccon1} \\
&& \cD^{ia}{\cal V}=\frac{1}{2}\cbD{}^i_b{\cal V}^{ab},\quad
\cbD{}^{ia}{\cal V}=\frac{1}{2}\cD^i_b{\cal V}^{ab}\;. \label{6ccon2}
\eea
The constraints \p{6ccon1} define ${\cal V}^{ab}$ as the $N{=}8$
superfield encompassing the off-shell multiplet ({\bf 3, 8, 5}),
while, as one can deduce from \p{6ccon1}, \p{6ccon2}, the $N{=}8$
superfield ${\cal V}$ has the content ({\bf 1, 8, 7}). Then the constraints
\p{6ccon2} establish relations between the fermions in these two superfields
and reduce the number of independent auxiliary fields to four, so that
we end up, once again, with the irreducible $N{=}8$ multiplet ({\bf 4, 8, 4}).

Two sets of $N{=}4$ covariant derivatives
$$\left( D^a,\bD{}^a\right) \equiv
\left( \cD^{1a},\cbD^{2a}\right) \mbox{ and  }
\left( \bar\nabla{}^a,\nabla^a\right) \equiv \left( \cD^{2a},\cbD^{1a}\right)$$
 are naturally realized in terms of the
$N{=}4$ superspaces
$\left( t,\theta_{1a}+i\vt_{1a},\theta_{2a}-i\vt_{2a}\right)$ and
$\left( t, \theta_{2a}+i\vt_{2a}\right. ,$ $\left.
\theta_{1a}-i\vt_{1a}\right)$.
In terms of the new derivatives, the constraints \p{6ccon1}, \p{6ccon2}
become
\bea
&& D^{(a}\cV^{bc)}=\bD{}^{(a}\cV^{bc)}
=\nabla^{(a}\cV^{bc)}=\bar\nabla{}^{(a}\cV^{bc)}=0 , \nn
&& D^a \cV=\frac{1}{2}\nabla_b \cV^{ab},\; \bD{}^a \cV
=\frac{1}{2}\bar\nabla_b \cV^{ab},\quad
\nabla^a \cV=\frac{1}{2}D_b \cV^{ab},\;
\bar\nabla{}^a \cV=\frac{1}{2}\bD_b \cV^{ab}.\label{6ccon3}
\eea
Now we see that the $\nabla^a, \bar\nabla_a$ derivatives of the
superfields $\cV, \cV^{ab}$ are expressed as $D^a,\bD{}^a$
of the superfields $\cV^{ab},\cV\,$, respectively. Thus, in the
$\left( \theta_{2a}+i\vt_{2a},\theta_{1a}-i\vt_{1a}\right)$
expansion of the superfields
$\cV,\cV^{ab}$ only the first components (i.e. those of zero order in
the coordinates $\left(\theta_{2a}+i\vt_{2a},
\theta_{1a}-i\vt_{1a}\right)$) are independent $N{=}4$ superfields.
We denote them  $v, v^{ab}$. The hidden $N{=}4$ supersymmetry is realized
on these $N{=}4$ superfields as
\be\label{6ctr1}
\delta v =-\frac{1}{2} \eta_a D_b v^{ab}+\frac{1}{2}\bar\eta_a\bD_b v^{ab},\;
\delta v^{ab}=\frac{4}{3}\left( \eta^{(a}D^{b)}v
-\bar\eta{}^{(a}\bD{}^{b)}v\right),
\ee
while the superfields themselves obey the constraints
\be\label{6ccon4}
D^{(a}v^{bc)}=\bD{}^{(a}v^{bc)}=0,\quad D^{(a}\bD{}^{b)}v=0\;,
\ee
which are remnant of the $N{=}8$ superfield constraints \p{6ccon3}.

The invariant free action reads
\be\label{6caction}
S=\int dt d^4\theta \left( v^2 -\frac{3}{8} v^{ab}v_{ab}\right).
\ee
\vspace{0.5cm}

\noindent{\bf 3. } ({\bf 4, 8, 4}) = ({\bf 2, 4, 2}) $\oplus$ ({\bf 2, 4, 2})\\

This case is a little bit more tricky. First of all, we define the
new set of $N{=}8$ superfields $\cW,\Phi$ in terms of $\cV^{ij},\cV$
defined earlier in \p{6csf}
\be\label{6bsf}
\cW \equiv \cV^{11}, \overline\cW\equiv \cV^{22},
\quad \Phi\equiv\frac{2}{3}\left( \cV+\frac{3}{2}\cV^{12}\right),\;
\overline\Phi\equiv\frac{2}{3}\left( \cV-\frac{3}{2}\cV^{12}\right)
\ee
and construct two new sets of $N{=}4$ derivatives $D^i,\nabla^i$
{}from those defined in \p{4bder}
\bea\label{6bder}
&&D^i=\frac{1}{\sqrt{2}}\left(\cD^{i1}+\cbD{}^{i1}\right),\;
\bD{}^i=\frac{1}{\sqrt{2}}\left(\cD^{i2}+\cbD{}^{i2}\right),\nn
&& \nabla^i=\frac{1}{\sqrt{2}}\left(\cD^{i1}-\cbD{}^{i1}\right),\;
\overline\nabla{}^i=-\frac{1}{\sqrt{2}}\left(\cD^{i2} - \cbD{}^{i2}\right).\;
\eea

The basic constraints \p{6ccon1}, \p{6ccon2} can be rewritten in terms
of the superfields $\cW,\Phi$ and the derivatives $D^i, \nabla^i$ as
\bea\label{6bcon1}
&& D^i \cW= \nabla^i \cW=0,\; \bD{}^i \overline\cW
=\overline\nabla{}^i \overline\cW=0,\quad   D^i \Phi
=\overline\nabla{}^i \Phi =0 ,\;
 \nabla^i \overline\Phi=\bD{}^i\overline\Phi=0,\nn
&& \bD{}^i \cW - D^i \overline\Phi=0,\;
D^i \overline\cW+\bD{}^i \Phi=0,\quad
\overline\nabla{}^i \cW-\nabla^i \Phi=0,\;
\nabla^i \overline\cW+\overline\nabla{}^i \overline\Phi=0.
\eea
The proper $N{=}4$ superspace is defined as the one on
which the covariant derivatives $D^1, \bD{}^2, \nabla^1,
\overline\nabla{}^2$
are naturally realized. The constraints \p{6bcon1} imply
that the remaining set of covariant derivatives, i.e.
$D^2, \bD{}^1, \nabla^2, \overline\nabla{}^1$,
when acting on every involved $N{=}8$ superfield, can be expressed
as spinor derivatives from the the first set acting on some
other $N{=}8$ superfield. Thus the first $N{=}4$
superfield components of the
$N{=}8$ superfields $\cW, \Phi$ are the only independent $N{=}4$
superfield projections. The transformations of the implicit
$N{=}4$ Poincar\'e supersymmetry have the following form in terms of
these $N{=}4$ superfields:
\bea\label{6btransf}
&& \delta w = \beps D^1 \bar\phi +
\bar\eta \nabla^1 \phi,\quad \delta \phi
= -\eta\overline\nabla_1 w -\beps D^1 {\bar w}\;, \nn
&& \delta {\bar w} = \epsilon \bD_1 \phi
+ \eta \overline\nabla_1{\bar\phi},\quad
\delta{\bar\phi}=-\epsilon\bD_1 w-\bar\eta \nabla^1 {\bar w}\;.
\eea

The free invariant action reads
\be\label{6baction}
S=\int dt d^4 \theta \left( w{\bar w}-\phi\bar\phi \right).
\ee

\subsection{Supermultiplet ({\bf 5, 8, 3})}
This supermultiplet has been considered in detail
in Refs. \cite{{DE},{bikl2}}. It was termed there the `$N{=}8$ vector multiplet'.
Here we sketch its main properties.

In order to describe this supermultiplet, one should introduce
five bosonic $N{=}8$ superfields ${\cal V}_{\alpha a},{\cal U}$
obeying the constraints
\be\label{5constr}
D^{ib}{\cal V}_{\alpha a} + \delta_a^b \nabla_\alpha^i{\cal U}=0\;, \quad
\nabla^{\beta i}{\cal V}_{\alpha a} + \delta_\alpha^\beta D^i_a {\cal U}=0
\;.
\ee

It is worth noting that the constraints \p{5constr} are covariant
not only under three $SU(2)$ automorphism groups (realized on the
doublet indices $i$, $a$ and $\alpha$), but also under the $SO(5)$
automorphisms. These $SO(5)$ transformations mix the spinor derivatives
$D^{ia}$ and $\nabla^{\alpha i}$ in the indices $\alpha$ and $a$,
while two $SU(2)$ groups realized on these indices constitute $SO(4)\subset SO(5)$.
The superfields $\cU,\cV^{\alpha a}$ form an $SO(5)$ vector:
under $SO(5)$ transformations belonging to the coset $SO(5)/SO(4)$
they transform  as
\be\label{so5a}
\delta \cV_{\alpha a}= a_{\alpha a}\; \cU \;,
\quad \delta \cU=-2 a_{\alpha a}\; \cV^{\alpha a}\,.
\ee

As in the previous cases we may consider two different splittings
of the $N{=}8$ vector multiplet into irreducible $N{=}4$ superfields
\begin{itemize}
\item {\bf 1. } ({\bf 5, 8, 3}) = ({\bf 1, 4, 3}) $\oplus$ ({\bf 4, 4, 0})
\item {\bf 2. } ({\bf 5, 8, 3}) = ({\bf 3, 4, 1}) $\oplus$ ({\bf 2, 4, 2})
\end{itemize}
Once again, they correspond to two different choices of
the $N{=}4, d{=}1$ superspace as a subspace in the original $N{=}8, d{=}1$ superspace.
\vspace{0.5cm}

\noindent{\bf 1. } ({\bf 5, 8, 3}) = ({\bf 1, 4, 3}) $\oplus$ ({\bf 4, 4, 0})\\

The relevant $N{=}4$ superspace is $\mathbb{R}^{(1|4)}$ parameterized by the coordinates
$\left( t, \theta_{ia} \right)$ and defined in \p{ss1}. As in the previous cases,
it follows from the constraints \p{5constr} that the spinor derivatives of
all involved superfields with respect to $\vartheta_{i\alpha}$
are expressed in terms of spinor derivatives with respect to
$\theta_{i a}$. Thus the only essential $N{=}4$ superfield components
of $\cV_{\alpha a}$ and $\cU$ in their $\vartheta$-expansion are the
first ones
\be\label{n4comp}
v_{\alpha a} \equiv \cV_{\alpha a}|_{\vt=0}\;,\quad u \equiv \cU|_{\vt=0}\,.
\ee
They accommodate the whole off-shell component content of
the $N{=}8$ vector multiplet. These five bosonic $N{=}4$ superfields are
subjected, in virtue of \p{5constr}, to the irreducibility constraints
in $N{=}4$ superspace
\be\label{5constra}
D^{i(a} v^{b)\alpha}=0,\quad D^{i(a}D_i^{b)} u= 0.
\ee
Thus, from the $N{=}4$ superspace standpoint, the vector $N{=}8$ supermultiplet
is the sum of the $N{=}4,d{=}1$ hypermultiplet $ v_{\alpha a}$ with
the  off-shell component contents ({\bf 4, 4, 0}) and the
$N{=}4$ `old' tensor multiplet $u$ with the contents ({\bf 1, 4, 3}).

The transformations of the implicit $N{=}4$ Poincar\'e supersymmetry read
\be\label{5n4transfa}
\delta v_{a\alpha}= \eta_{i\alpha} D^i_a u\;, \quad \delta u
={1\over 2}\eta_{i\alpha}D^{ia} v_a^\alpha\;.
\ee
\vspace{0.5cm}

\noindent{\bf 2. } ({\bf 5, 8, 3}) = ({\bf 3, 4, 1}) $\oplus$ ({\bf 2, 4, 1})\\

Another interesting $N{=}4$ superfield splitting of the $N{=}8$
vector multiplet can be achieved by passing to the complex parametrization
of the $N{=}8$ superspace as
$$
\left(t,\Theta_{i\alpha}= \theta_{i\alpha}+i\vartheta_{\alpha i},
\bar\Theta^{i\alpha}= \theta^{i\alpha} -i\vartheta^{\alpha i}\right)
$$
where we have identified the indices $a$ and $\alpha$, thus reducing the number of
manifest $SU(2)$ automorphism symmetries to just two. In this superspace
the covariant derivatives $\cD^{i\alpha},\cbD{}^{j\beta}$
defined in \p{4bder} (with the identification of indices just mentioned)
are naturally realized. We are also led to define new superfields
\bea
\cV \equiv -\epsilon_{\alpha a}\cV^{\alpha a}\,,
\quad \cW^{\alpha\beta} \equiv \cV^{(\alpha \beta)} = {1\over 2}
\left( \cV^{\alpha \beta}+\cV^{\beta \alpha}\right)\,,
\quad \cW \equiv \cV +i{\cal U}\,,
\quad \overline{\cW} \equiv \cV - i{\cal U}\,.
\eea

In this basis of $N{=}8$ superspace the original constraints
\p{5constr} amount to
\bea\label{5constrb}
&&\cD^{i\alpha}\cW^{\beta\gamma}=
-\frac{1}{4}\left( \epsilon^{\beta\alpha}\cbD{}^{i\gamma}\overline{\cW}+
\epsilon^{\gamma\alpha}\cbD{}^{i\beta}\overline{\cW}\right),\;
\cbD{}^{i\alpha}\cW^{\beta\gamma}=-\frac{1}{4}\left( \epsilon^{\beta\alpha}
\cD^{i\gamma} \cW+\epsilon^{\gamma\alpha}\cD{}^{i\beta}\cW\right),\nn
&& \cD^{i\alpha}\overline{\cW}=0,\;\cbD^{i\alpha} \cW=0\,,
\quad (\cD^{k\alpha}\cD^i_\alpha) \cW
=(\cbD^{k}_{\alpha}\cbD^{i\alpha})\overline{\cW}\,.
\eea
Next, we single out the $N{=}4, d{=}1$ superspace  as
$\left( t, \theta_\alpha \equiv \Theta_{1\alpha}, \bar\theta^{\alpha}\right)$
and split our $N{=}8$ superfields into the $N{=}4$ ones in the standard way.
As in all previous cases,  the spinor derivatives of
each $N{=}8$ superfield with respect to
$\overline{\Theta}^{2\alpha}$ and $\Theta_{2\alpha}$, as
a consequence of the constraints \p{5constrb}, are expressed as derivatives
of some other superfields with respect to $\bar\theta^{\alpha}$
and $\theta_{\alpha}$. Therefore, only the first
(i.e. taken at $\overline{\Theta}^{2a}=0$ and $\Theta_{2a}=0$)
$N{=}4$ superfield components of the $N{=}8$ superfields really matter.
They accommodate the entire off-shell field content of the multiplet.
These $N{=}4$ superfields are defined as
\be
\left. \phi \equiv \cW \right|\,,
\quad, \left. \bar\phi \equiv \overline{\cW}\right|\,,
\quad \left. w^{\alpha\beta} \equiv \cW^{\alpha\beta}\right|
\ee
and satisfy the constraints following from \p{5constrb}
\be\label{5finalconstr}
{\cD}^{\alpha} \bar\phi =0\,,\quad \cbD_{\alpha}\phi=0\,,\quad
\cD^{(\alpha}w^{\beta\gamma)}=\cbD{}^{(\alpha}w^{\beta\gamma)}=0\,,
\quad \cD^\alpha
\equiv \cD^{1\alpha}\,, \;\cbD_\alpha \equiv \cbD_{1\alpha}\,.
\ee
They tell us that the $N{=}4$ superfields $\phi$ and $\bar\phi$
form the standard $N{=}4$ chiral multiplet ({\bf 2, 4, 2}),
while the $N{=}4$ superfield $w^{\alpha\beta}$ represents the $N{=}4$
tensor multiplet ({\bf 3, 4, 1}).

The implicit $N{=}4$ supersymmetry is realized on $w^{\alpha\beta}$,
$\phi$ and $\bar\phi$ as
\be\label{6tra}
\delta w^{\alpha\beta}=\frac{1}{2}\left(\eta^{(\alpha}\cbD^{\beta)}\bar\phi
-\bar\eta{}^{(\alpha}\cD{}^{\beta)}\phi\right)\,,\quad \delta \phi=
\frac{4}{3}\eta_\alpha \cbD^\beta w_\beta^\alpha\,,\quad
\delta  \bar\phi=-\frac{4}{3}\bar\eta{}^{\alpha}\cD_\beta w_\alpha^\beta\,.
\ee

An analysis of $N{=}8$ supersymmetric actions for the $N{=}8$ vector multiplet
may be found in \cite{bikl2}.

\subsection{Supermultiplet ({\bf 6, 8, 2})}
This supermultiplet can be described by two $N{=}8$ tensor
multiplets $\cV^{ij}$ and $\cW^{ab}\,$,
\be\label{7con1}
D^{(i}_a\cV^{jk)}=0\,,\; \nabla^{(i}_a\cV^{jk)}=0\,,\quad
D^{(a}_i\cW^{bc)}=0\,,\; \nabla^{(a}_i\cW^{bc)}=0\,,
\ee
with the additional constraints
\be\label{7con2}
D^a_j \cV^{ij}=\nabla^{bi}\cW^a_b,\quad \nabla^a_j \cV^{ij}=-D^i_b \cW^{ab} \;.
\ee
The role of the latter constraints is to identify the eight fermions,
which are present in $\cV^{ij}$, with the fermions from
$\cW^{ab}$, and to reduce the number of independent auxiliary fields
in both superfields to two
\be\label{7auxil}
F_1=D^a_iD_{aj}\cV^{ij}|\,,\quad F_2 = D^i_a D_{ib}\cW^{ab}|\,,
\ee
where $|$ means here restriction to $\theta_{ia}=\vt_{ia}=0\,$.

There are two different possibilities to split this $N{=}8$ multiplet
into $N{=}4$ ones

\begin{itemize}
\item {\bf 1. } ({\bf 6, 8, 2}) = ({\bf 3, 4, 1}) $\oplus$ ({\bf 3, 4, 1})
\item {\bf 2. } ({\bf 6, 8, 2}) = ({\bf 4, 4, 0}) $\oplus$ ({\bf 2, 4, 2})
\end{itemize}

As before, we discuss the peculiarities of both decompositions.
\vspace{0.5cm}

\noindent{\bf 1. } ({\bf 6, 8, 2}) = ({\bf 3, 4, 1}) $\oplus$ ({\bf 3, 4, 1})\\

The corresponding $N{=}4$ supersubspace is \p{ss1}.
The $N{=}8$ constraints imply that the only
essential $N{=}4$ superfield components
of $\cV^{ij}$ and $\cW^{ab}$ in their $\vartheta$-expansion are the first ones
\be\label{7comp1}
v^{ij} \equiv \cV^{ij}|\;,\quad w^{ab} \equiv \cW^{ab}| \,.
\ee
These six bosonic $N{=}4$ superfields are
subjected, in virtue of \p{7con1}, \p{7con2}, to the irreducibility
constraints in $N{=}4$ superspace
\be\label{7con3}
D^{a(i} v^{jk)}=0\,,\quad D^{i(a} w^{bc)}= 0\,.
\ee
Thus, the  $N{=}8$ supermultiplet ({\bf 6, 8, 2}) amounts to the sum of
two $N{=}4,d{=}1$ tensor multiplets $v^{ij}, w^{ab}$ with
the off-shell field contents ({\bf 3, 4, 1}) $\oplus$ ({\bf 3, 4, 1}).

The transformations of the implicit $N{=}4$ Poincar\'e supersymmetry
are
\be\label{7transf1}
\delta v^{ij}=-\frac{2}{3}\eta^{(i}_a D^{j)}_b w^{ab},\quad
\delta w^{ab}=\frac{2}{3} \eta^{(a}_i D^{b)}_j v^{ij}.
\ee

The free $N{=}8$ supersymmetric action has the following form:
\be\label{7action1}
S=\int dt d^4 \theta \left( v^2- w^2 \right).
\ee
\vspace{0.5cm}

\noindent{\bf 2. } ({\bf 6, 8, 2}) = ({\bf 4, 4, 0}) $\oplus$ ({\bf 2, 4, 2})\\

In this case, to describe the ({\bf 6, 8, 2}) multiplet,
we combine two $N{=}4$ superfields:, i.e. the chiral
superfield
\be\label{7cona}
D^i \phi=\bD{}^i \bar\phi=0
\ee
and the hypermultiplet $q^{ia}$
\be\label{7cona1}
D^{(i} q^{j)a}=\bD{}^{(i} q^{j)a}=0 \:.
\ee

The transformations of the implicit $N{=}4$ supersymmetry read
\be\label{7tra}
\delta q^{ia}=\beps{}^a D^i\bar\phi +\epsilon^a \bD{}^i \phi\;, \quad
\delta\phi=-\frac{1}{2}\beps{}^a D^i q_{ia}\;, \;
\delta\bar\phi=-\frac{1}{2}\epsilon^a\bD{}^i q_{ia}\;.
\ee

The invariant free action reads
\be\label{7actiona1}
S_{free}=\int dt d^4\theta \left( q^2-4\phi\bar\phi\right)\;.
\ee

\subsection{Supermultiplet ({\bf 7, 8, 1})}
This supermultiplet has a natural description in terms of
two $N{=}8$ superfields $\cV^{ij}$ and $\cQ^{a\alpha}$ satisfying
the constraints
\bea
&& D^{(i a}\cV^{jk)}=0,\quad \nabla^{\alpha(i}\cV^{jk)}=0, \quad
 D^{i(a}\cQ^{\alpha b)}=0, \quad \nabla^{(\alpha}_i\cQ^{\beta)}_a=0,
\label{8con1} \\
&& D_j^a \cV^{ij}=i\nabla^i_\alpha \cQ^{a\alpha},
\quad \nabla^\alpha_j \cV^{ij}=-iD^i_a \cQ^{a\alpha} .\label{8con2}
\eea
The constraints \p{8con1} leave in the superfields $\cV^{ij}$
and $\cQ^{a\alpha}$ the sets ({\bf 3, 8, 5}) and $({\bf 4,\, 8,\, 4})$
of irreducible components, respectively. The role of the
constraints \p{8con2} is to identify the fermions in the
superfields $\cV^{ij}$ and $\cQ^{a\alpha}$ and reduce the
total number of independent auxiliary components in both
superfields to just one.

For this supermultiplet there is a unique splitting into
$N{=}4$ superfields as
$$
({\bf 7,\, 8,\, 1}) = ({\bf 3,\, 4,\, 1}) \oplus ({\bf 4,\, 4,\, 0}).
$$
The proper $N{=}4$ superspace is parameterized by the coordinates
$\left( t, \theta_{ia} \right)$.
The constraints \p{8con1}, \p{8con2} imply that the only
essential $N{=}4$ superfield components in the  $\vartheta$-expansion
of $\cV^{ij}$ and $\cQ^{a\alpha}$ are the first ones
\be\label{8comp}
v^{ij} \equiv \cV^{ij}|_{\vt=0}\;,\quad
q^{a\alpha} \equiv \cQ^{a\alpha}|_{\vt=0}\,.
\ee
These seven bosonic $N{=}4$ superfields are subjected, as a corollary of
\p{8con1}, \p{8con2}, to the irreducibility constraints in $N{=}4$ superspace
\be\label{8const}
D^{a(i} v^{jk)}=0\,,\quad D^{i(a} q^{b)\alpha} = 0\,.
\ee
Thus the $N{=}8$ supermultiplet ({\bf 7, 8, 1}) amounts to the sum of the
$N{=}4, d{=}1$ hypermultiplet $ q^{a\alpha }$ with
the ({\bf 4, 4, 0}) off-shell field content and the
$N{=}4$ tensor multiplet $v^{ij}$ with the ({\bf 3, 4, 1}) content.

The implicit $N{=}4$ Poincar\'e supersymmetry is realized by the
transformations
\be\label{8transfa}
\delta v^{ij}= -\frac{2i}{3}\,\eta^{(i}_{\alpha} D^{j)}_a q^{a\alpha}\;, \quad
\delta q^{a\alpha}
=-{i\over 2}\eta^{i\alpha}D^{ja} v_{ij}\;.
\ee

The free action can be also easily written
\be\label{8action}
S=\int dt d^4\theta \left[ v^2 -\frac{4}{3} q^2  \right].
\ee

\subsection{Supermultiplet ({\bf 8, 8, 0})}

This supermultiplet is analogous to the supermultiplet ({\bf 0, 8, 8}):
they differ in their overall Grassmann parity. It is described by the two
real bosonic $N{=}8$ superfields
$\cQ^{aA}, \Phi^{i\alpha}$ subjected to the constraints
\bea
&& D^{(i a}\Phi^{j)\alpha}=0\,,\; D^{i(a} \cQ^{b)A}=0\,,\quad
\nabla^{(\alpha A}\Phi^{\beta)}_i=0\,,\; \nabla^{\alpha( A} \cQ^{a B)}=0\,,
\label{9con1} \\
&& \nabla^{\alpha A} \cQ_A^a=-D^{ia}\Phi_i^\alpha\,,\quad
\nabla^{\alpha A}\Phi_\alpha^i=D^{ia} \cQ_a^A\,. \label{9con2}
\eea

Analogously to the case of the supermultiplet ({\bf 8, 8, 0}),
{}from the constraints \p{9con2} it follows that the spinor derivatives of
all involved superfields with respect to $\vartheta_{\alpha A}$
are expressed in terms of spinor derivatives with respect to
$\theta_{i a}$. Thus the only essential $N{=}4$ superfield components
in the $\vartheta$-expansion
of $\cQ^{aA}$ and $\Phi^{i\alpha}$ are the first ones
\be\label{9comp1}
q^{aA} \equiv \cQ^{aA}|_{\vt=0}\;,
\quad \phi^{i\alpha} \equiv \Phi^{i\alpha}|_{\vt=0}\,.
\ee
They accommodate the whole off-shell component content of the
multiplet ({\bf 8, 8, 0}). These  bosonic $N{=}4$ superfields are
subjected, as a consequence of \p{9con1}, \p{9con2},
to the irreducibility constraints in $N{=}4$ superspace
\be\label{9con3}
D^{a(i}\phi^{j)\alpha}=0\,,\; D^{i(a}q^{b)A}=0\,.
\ee
Thus the  $N{=}8$ supermultiplet ({\bf 8, 8, 0}) can be represented
as the sum of two $N{=}4, d{=}1$  hypermultiplets  with
the off-shell component contents ({\bf 4, 4, 0}) $\oplus$ ({\bf 4, 4, 0}).

The transformations of the implicit $N{=}4$ Poincar\'e supersymmetry
in this last case are as follows:
\be\label{9tr1}
\delta q^{aA}=-\frac{1}{2}\eta^{A\alpha} D^{ia} \phi_{i\alpha},\quad
\delta \phi^{i\alpha}=\frac{1}{2}\eta^\alpha_A D^i_a q^{aA}.
\ee

The invariant free action reads
\be\label{9action1}
S=\int dt d^4\theta \left[ q^2 - \phi^2\right].
\ee

The most general action still respecting four $SU(2)$ automorphism symmetries
has the following form:
\be\label{9action2}
S=\int dt d^4\theta F( q^2, \phi^2),
\ee
where, as a necessary condition of $N{=}8$ supersymmetry,
the function $F( q^2, \phi^2)$ should obey the equation
\be\label{9action2a}
\frac{\partial^2}{\partial q^2 \partial q^2} \left( q^2 F( q^2, \phi^2)\right)+
\frac{\partial^2}{\partial \phi^2 \partial \phi^2}
\left( \phi^2 F( q^2,\phi^2)\right) = 0.
\ee

Thus, we presented superfield formulations
of the full amount of off-shell $N{=}8, d{=}1$ supermultiplets
with {\bf 8} physical fermions, both in $N{=}8$ and $N{=}4$ superspaces.
We listed all possible $N{=}4$ superfield splittings of these multiplets.

\section{Summary and conclusions}
In these Lectures we reviewed the superfield approach to extended supersymmetric
one-dimensional models. We presented superfield formulations
of the full amount of off-shell $N=4$ and $N{=}8, d{=}1$ supermultiplets
with {\bf 4} and {\bf 8} physical fermions, respectively.
We also demonstrated how to reproduce $N=4$ supermultiplets
{}from nonlinear realizations of the $N{=}4, d{=}1$ superconformal group.

It should be pointed out that here we addressed only those
multiplets which satisfy linear constraints in superspace. As we know,
there exist $N{=}4, d{=}1$ multiplets with nonlinear defining constraints
(e.g. nonlinear versions of the chiral ({\bf 2, 4, 2}) multiplet,
as well as of the hypermultiplet ({\bf 4, 4, 0}).
It would be interesting to construct analogous nonlinear versions of some
$N{=}8$ multiplets from the above set. Moreover, for all our linear supermultiplets
the bosonic metrics of the general sigma-model type actions are proven to be
conformally flat. This immediately raises the question - how to describe $N=4$ and $N=8, d=1$
sigma models with hyper-K\"{a}hler metrics in the target space? For the $N=8$ cases it seems the
unique possibility is to use {\it infinite} dimensional supermultiplets, as in the
case of $N=2, d=4$ supersymmetry \cite{harm,book}. But for $N=4, d=1$ supersymmetric models
infinite dimensional supermultiplets do not exist! Therefore, the unique possibility in this
case is to use some nonlinear supermultiplets. In this respect, the harmonic superspace
approach seems to yield the most relevant framework. So, all results we discussed should be regarded as
preparatory for a more detailed study of $N{=}4, 8$ $d=1$ supersymmetric models.

\section*{Acknowledgements}
We thank to E.~Ivanov and O.~Lechtenfeld in collaboration with whom some of the presented
results where obtained. Useful conversations with A. Nersessian are acknowledged
with pleasure.
This research was partially supported by the
the INTAS-00-00254  grant,
RFBR-DFG grant No 02-02-04002, grant DFG No 436 RUS 113/669, and RFBR grant
No 03-02-17440.
S.K. thanks INFN - Laboratori Nazionali di Frascati  for the warm
hospitality extended to him during the course of this work.

%
%

%
%



\printindex
\end{document}